\documentclass[aps,pra,12pt,onecolumn,superscriptaddress]{revtex4-1}

\usepackage{amsmath,amsfonts,amssymb,color,epsfig,graphics,graphicx,latexsym,revsymb,theorem,url,verbatim,epstopdf}

\usepackage{amssymb}
\usepackage{color}
\usepackage{amsmath,epic,curves,amscd}
\usepackage[english]{babel}
\usepackage{graphicx}
\usepackage{comment}
\usepackage{appendix}
\usepackage{mathdots}

\usepackage{hyperref}

\newtheorem{definition}{Definition}
\newtheorem{proposition}[definition]{Proposition}
\newtheorem{lemma}[definition]{Lemma}

\newtheorem{theorem}[definition]{Theorem}
\newtheorem{corollary}[definition]{Corollary}
\newtheorem{conjecture}[definition]{Conjecture}

\newtheorem{remark}[definition]{Remark}
\newtheorem{example}[definition]{Example}
\newtheorem{question}[definition]{Question}

\def\squareforqed{\hbox{\rlap{$\sqcap$}$\sqcup$}}
\def\qed{\ifmmode\squareforqed\else{\unskip\nobreak\hfil
\penalty50\hskip1em\null\nobreak\hfil\squareforqed
\parfillskip=0pt\finalhyphendemerits=0\endgraf}\fi}
\def\endenv{\ifmmode\;\else{\unskip\nobreak\hfil
\penalty50\hskip1em\null\nobreak\hfil\;
\parfillskip=0pt\finalhyphendemerits=0\endgraf}\fi}
\newenvironment{proof}{\noindent \textbf{{Proof.~} }}{\qed}
\def\Dbar{\leavevmode\lower.6ex\hbox to 0pt
{\hskip-.23ex\accent"16\hss}D}
\makeatletter
\def\url@leostyle{%
  \@ifundefined{selectfont}{\def\UrlFont{\sf}}{\def\UrlFont{\small\ttfamily}}}
\makeatother
\def\bcj{\begin{conjecture}}
\def\ecj{\end{conjecture}}
\def\bcr{\begin{corollary}}
\def\ecr{\end{corollary}}
\def\bd{\begin{definition}}
\def\ed{\end{definition}}
\def\bea{\begin{eqnarray}}
\def\eea{\end{eqnarray}}
\def\bem{\begin{enumerate}}
\def\eem{\end{enumerate}}
\def\bex{\begin{example}}
\def\eex{\end{example}}
\def\bim{\begin{itemize}}
\def\eim{\end{itemize}}
\def\bl{\begin{lemma}}
\def\el{\end{lemma}}
\def\bpf{\begin{proof}}
\def\epf{\end{proof}}
\def\bpp{\begin{proposition}}
\def\epp{\end{proposition}}
\def\bqu{\begin{question}}
\def\equ{\end{question}}
\def\br{\begin{remark}}
\def\er{\end{remark}}
\def\bt{\begin{theorem}}
\def\et{\end{theorem}}

\def\btb{\begin{tabular}}
\def\etb{\end{tabular}}

\newcommand{\nc}{\newcommand}

\def\a{\alpha}
\def\b{\beta}

\def\m{\mu}
\def\n{\nu}

\def\r{\rho}

 \nc{\bbA}{\mathbb{A}} \nc{\bbB}{\mathbb{B}} \nc{\bbC}{\mathbb{C}}
 \nc{\bbD}{\mathbb{D}} \nc{\bbE}{\mathbb{E}} \nc{\bbF}{\mathbb{F}}
 \nc{\bbG}{\mathbb{G}} \nc{\bbH}{\mathbb{H}} \nc{\bbI}{\mathbb{I}}
 \nc{\bbJ}{\mathbb{J}} \nc{\bbK}{\mathbb{K}} \nc{\bbL}{\mathbb{L}}
 \nc{\bbM}{\mathbb{M}} \nc{\bbN}{\mathbb{N}} \nc{\bbO}{\mathbb{O}}
 \nc{\bbP}{\mathbb{P}} \nc{\bbQ}{\mathbb{Q}} \nc{\bbR}{\mathbb{R}}
 \nc{\bbS}{\mathbb{S}} \nc{\bbT}{\mathbb{T}} \nc{\bbU}{\mathbb{U}}
 \nc{\bbV}{\mathbb{V}} \nc{\bbW}{\mathbb{W}} \nc{\bbX}{\mathbb{X}}
 \nc{\bbZ}{\mathbb{Z}}

 \nc{\bA}{{\bf A}} \nc{\bB}{{\bf B}} \nc{\bC}{{\bf C}}
 \nc{\bD}{{\bf D}} \nc{\bE}{{\bf E}} \nc{\bF}{{\bf F}}
 \nc{\bG}{{\bf G}} \nc{\bH}{{\bf H}} \nc{\bI}{{\bf I}}
 \nc{\bJ}{{\bf J}} \nc{\bK}{{\bf K}} \nc{\bL}{{\bf L}}
 \nc{\bM}{{\bf M}} \nc{\bN}{{\bf N}} \nc{\bO}{{\bf O}}
 \nc{\bP}{{\bf P}} \nc{\bQ}{{\bf Q}} \nc{\bR}{{\bf R}}
 \nc{\bS}{{\bf S}} \nc{\bT}{{\bf T}} \nc{\bU}{{\bf U}}
 \nc{\bV}{{\bf V}} \nc{\bW}{{\bf W}} \nc{\bX}{{\bf X}}
 \nc{\bZ}{{\bf Z}}

\nc{\cA}{{\cal A}} \nc{\cB}{{\cal B}} \nc{\cC}{{\cal C}}
\nc{\cD}{{\cal D}} \nc{\cE}{{\cal E}} \nc{\cF}{{\cal F}}
\nc{\cG}{{\cal G}} \nc{\cH}{{\cal H}} \nc{\cI}{{\cal I}}
\nc{\cJ}{{\cal J}} \nc{\cK}{{\cal K}} \nc{\cL}{{\cal L}}
\nc{\cM}{{\cal M}} \nc{\cN}{{\cal N}} \nc{\cO}{{\cal O}}
\nc{\cP}{{\cal P}} \nc{\cQ}{{\cal Q}} \nc{\cR}{{\cal R}}
\nc{\cS}{{\cal S}} \nc{\cT}{{\cal T}} \nc{\cU}{{\cal U}}
\nc{\cV}{{\cal V}} \nc{\cW}{{\cal W}} \nc{\cX}{{\cal X}}
\nc{\cZ}{{\cal Z}}

\nc{\hA}{{\hat{A}}} \nc{\hB}{{\hat{B}}} \nc{\hC}{{\hat{C}}}
\nc{\hD}{{\hat{D}}} \nc{\hE}{{\hat{E}}} \nc{\hF}{{\hat{F}}}
\nc{\hG}{{\hat{G}}} \nc{\hH}{{\hat{H}}} \nc{\hI}{{\hat{I}}}
\nc{\hJ}{{\hat{J}}} \nc{\hK}{{\hat{K}}} \nc{\hL}{{\hat{L}}}
\nc{\hM}{{\hat{M}}} \nc{\hN}{{\hat{N}}} \nc{\hO}{{\hat{O}}}
\nc{\hP}{{\hat{P}}} \nc{\hR}{{\hat{R}}} \nc{\hS}{{\hat{S}}}
\nc{\hT}{{\hat{T}}} \nc{\hU}{{\hat{U}}} \nc{\hV}{{\hat{V}}}
\nc{\hW}{{\hat{W}}} \nc{\hX}{{\hat{X}}} \nc{\hZ}{{\hat{Z}}}

\nc{\hn}{{\hat{n}}}

\def\dim{\mathop{\rm Dim}}

\def\tr{\mathop{\rm Tr}}

\def\Un{{\mbox{\rm U}}}

\def\ox{\otimes}

\def\ra{\rightarrow}

\newcommand{\ket}[1]{|#1\rangle}
\newcommand{\proj}[1]{| #1\rangle\!\langle #1 |}

\newcommand{\braket}[2]{\langle#1|#2\rangle}

\newcommand{\cmp}{Commun. Math. Phys.}

\newcommand{\jmp}{J. Math. Phys.}

\begin{document}
\title{Orthogonal product bases of four qubits}

\author{Lin Chen}
\email{linchen@buaa.edu.cn (corresponding author)}
\affiliation{School of Mathematics and Systems Science, Beihang University, Beijing 100191, China}
\affiliation{International Research Institute for Multidisciplinary Science, Beihang University, Beijing 100191, China}

\def\Dbar{\leavevmode\lower.6ex\hbox to 0pt
{\hskip-.23ex\accent"16\hss}D}
\author {{ Dragomir {\v{Z} \Dbar}okovi{\'c}}}
\email{djokovic@uwaterloo.ca}
\affiliation{Department of Pure Mathematics and Institute for
Quantum Computing, University of Waterloo, Waterloo, Ontario, N2L
3G1, Canada}

\begin{abstract}
An orthogonal product basis (OPB) of a finite-dimensional Hilbert 
space $\cH=\cH_1\otimes\cH_2\otimes\cdots\otimes\cH_n$ 
is an orthonormal basis of $\cH$ consisting of product vectors
$\ket{x_1}\otimes\ket{x_2}\otimes\cdots\otimes\ket{x_n}$.
We show that the problem of constructing the OPBs of an $n$-qubit system can be reduced to a purely combinatorial problem. 
We solve this combinatorial problem in the case of four qubits 
and obtain 33 multiparameter families of OPBs. 
Each OPB of four qubits is equivalent, under local unitary 
operations and qubit permutations, to an OPB belonging to at 
least one of these families.
\end{abstract}

\date{\today}


\maketitle

\tableofcontents

\section{Introduction}\label{sec1}

The local operations and classical communication (LOCC) are the fundamental measurements for many quantum-information protocols and problems \cite{clm14}. 
The quantum teleportation is carried out using LOCC and quantum entanglement \cite{bbc93}, the well-known distillability problem \cite{dss00,dcl00} and distinguishing of quantum states are investigated under LOCC \cite{dms03}.
The LOCC-indistinguishable product states imply the quantum nonlocality without entanglement \cite{bdf99}. It is known that 
the unextendible product bases (UPBs) \cite{bdf99} and irreducible orthogonal product bases (OPBs) 
\cite[Theorem 3]{fs09} are LOCC-indistinguishable. 
(For the definition of reducible and irreducible OPBs see 
section \ref{sec:opb}.) 
The irreducible three-qubit and two-qutrit OPBs have been classified in \cite{fs09}.

The difficulty of constructing and classifying multiqubit OPBs increases rapidly with the number of qubits. We show in this paper how the construction problem can be reduced to a purely combinatorial problem. 
In the case of four qubits, by solving the latter problem, we provide a method for the construction of all OPBs. 
The combinatorial problem deals with the special kind of 
$2^n\times n$ matrices $M\in\cO(n)$ where $n$ is the number of qubits (see Sec. \ref{sec:combinatorial} for the precise definition of $\cO(n)$.)
The entries of $M$ are formal variables $a,b,\ldots$, which represent unit vectors of various one-qubit Hilbert spaces 
$\cH_j$, $j=1,\ldots,n$. Each of these vector variables $a$ has as a companion another vector variable $a^\perp$. To a vector variable $a$ in column $j$ we may assign as values arbitrary unit vectors $\ket{\a}$ in $\cH_j$. Then to $a^\perp$ we have to assign the value $\ket{\a}^\perp\in\cH_j$. 
After assigning the unit vector values to each vector variable  of $M$, we obtain an OPB by simply taking the tensor product of the unit vectors in each row. Thus each $M$ determines an infinite family of OPBs which we denote by $\cF_M$.  A family $\cF_M$ is called maximal if it is not a proper subset of another family $\cF_N$. (This is equivalent to our formal 
definition \ref{def:maximal}.)

We say that two matrices are equivalent if the corresponding 
families of OPBs are equivalent by local unitary operations 
and qubit permutations. (Although our formal definition of 
equivalence of matrices $M$ is different, see Definition 
\ref{def:equivalent}, one can show that it is  equivalent to this one.)
The construction of OPBs reduces to the classification of maximal matrices $M$ up to equivalence. For convenience of the latter classification, we introduce weak equivalence in the set of maximal matrices and refer to weak equivalence classes as 
switching classes. This is the key notion for simplifying 
the enumeration of the equivalence classes in $\cO(n)$.
Each switching class is a union of ordinary equivalence classes.

There are infinitely many equivalence classes of OPBs of $n$ qubits (up to local unitary transformations and qubit permutations). On the other hand there are only finitely many equivalence classes of matrices in $\cO(n)$. Given an OPB, say 
$\cA$, after ordering its product vectors we can associate to it in a natural way a matrix $M\in\cO(n)$ such that $\cA\in\cF_M$ (see the end of section \ref{sec:combinatorial}). Thus the set of all OPBs is the union of finitely many families $\cF_M$.  Note that the matrix $M$ is not uniquely determined by $\cA$ as it depends on the choice of ordering of $\cA$ and naming of the vector variables when constructing $M$. However, all $M$s that we get in this way are equivalent to each other. Further, if $U$ is any local unitary operator then we have $U\cA\in\cF_M$. Hence, the equivalence class of $M$ depends only on the equivalence class of $\cA$. In other words, we obtain a well defined map from the set of equivalence classes of OPBs of $n$ qubits to the set of equivalence classes of $\cO(n)$. This mapping is onto but, of course, not one-to-one. 

To construct all OPBs, instead of arbitrary matrices $N\in\cO(n)$ one can use just the maximal ones. More precisely, let $M_1,M_2,\ldots,M_k$ be the list of representatives of the equivalence classes of maximal matrices in $\cO(n)$. Then any OPB $\cA$ is equivalent to one in some $\cF_{M_i}$. The main result of this paper (Theorem \ref{thm:4qubit}) provides such a list in the case of four qubits. However, we did not compute the list of all equivalence classes. (In principle such computation can be performed, say by a computer, but writing an efficient program is a challenge.) In the case of three qubits we computed the list of all equivalence classes (see Lemma \ref{le:3qubits}  
and its corollary).

In addition to constructing the OPBs we can also obtain a very coarse classification of them. Let $M\in\cO(n)$ be the matrix associated to $\cA$. If $\cA\in\cF_N$ for some $N\in\cO(n)$ then 
$\cF_M\subseteq\cF_N$. Thus, among all families $\cF_N$ 
containing $\cA$ there is the smallest one, namely $\cF_M$. 
In general, the condition $\cA\in\cF_N$ does not determine $N$ 
uniquely. A slight modification can correct this default. For that purpose we introduce the strict families $\cF^\#_M$. In order to define strict families we introduce a (non-transitive) binary relation, denoted by ``$<$'', on $\cO(n)$ (see section \ref{sec:combinatorial}). If $N<M$ then $\cF_N\subset\cF_M$. (If there is no $N$ such that $M<N$ then $M$ is maximal.) By definition, $\cF^\#_M$ consists of all OPBs in $\cF_M$ which do not belong to any $\cF_N$ with $N<M$. We still have that $\cA\in\cF^\#_M$. Moreover, $\cF^\#_M$ is the unique strict family containing $\cA$.

Let $\cR$ be a set of representatives of the equivalence classes 
of $\cO(n)$, a finite set. Our proposed classification is a partition of the set of all OPBs into finite number of classes, one for each $M\in\cR$. The class corresponding to $M\in\cR$ consists of all OPBs which are equivalent to an OPB belonging to the strict family $\cF^\#_M$.

In the case of three qubits, there exists exactly one reducible and one irreducible switching class of maximal matrices (see Lemma \ref{le:3qubits}). 
In the case of four qubits we prove in Theorem \ref{thm:4qubit} that there are in total 17 switching classes of maximal matrices, and we list their representatives (see \eqref{eq:sw1-3}-\eqref{eq:sw13-15}). 
Among these switching classes, 3 consist of reducible and 12 
of irreducible matrices. Two preliminary facts are proven in lemmas \ref{le:3q-2rows} and \ref{le:mu>5}. Due to its length, the rest of the proof of Theorem \ref{thm:4qubit} is given in a 
separate section, \ref{sec:proof}. When a switching class 
contains more then one equivalence class, we list their representatives in the appendix. In total, there are 33 equivalence classes of maximal matrices.

Throughout this paper we shall use the following notation.
Let $\cH=\cH_1\otimes\cH_2\otimes\cdots\otimes\cH_n$ be the
complex Hilbert space of a finite-dimensional $n$-partite quantum
system. We denote by $d_i$ the dimension of $\cH_i$, and so
$D:=\prod d_i$ is the dimension of $\cH$. To avoid trivial cases, we assume that each $d_i>1$ and $n>1$. 
A vector $\ket{x}\in\cH$ is {\em normalized} if $\|x\|=1$. We denote by $H$ the space of Hermitian operators $\r$ on $\cH$. Note that $H$ is a real vector space of dimension $D^2$. 
The mixed quantum states of this quantum system are represented
by their density matrices, i.e., operators $\r\in H$ which are
positive semidefinite $(\r\ge0)$ and have unit trace
$(\tr \r=1)$. 

We assume that an orthonormal (o. n.) basis is fixed in each $\cH_i$ and we use the standard notation 
$\ket{0},\ldots,\ket{d_i-1}$
for the corresponding basis vectors. 
A {\em product vector} is a nonzero vector of the form 
$\ket{x}=\ket{x_1}\ox\cdots\ox\ket{x_n}$ where 
$\ket{x_i}\in\cH_i$. We shall write this product vector also as 
$\ket{x_1,\ldots,x_n}$. 
When $\ket{x_1,\ldots,x_n}$ is a unit vector, we shall also 
assume that each $\ket{x_i}$ is a unit vector.
A {\em pure product state} is a state $\r$ of the form $\r=\proj{x}$ where $\ket{x}$ is a product vector. The product vectors 
$\ket{i_1,i_2,\ldots,i_n}$, $0\le i_k<d_k$, form an o. n. basis 
of $\cH$. We refer to this basis as the {\em standard basis}.

\section{Orthogonal product bases}
\label{sec:opb}

We say that an orthonormal basis of $\cH$ consisting of product 
vectors, 
\bea \label{eq:bazis-A}
\cA:=\{\ket{a_s}=\ket{a_{s,1},\ldots,a_{s,n}}:s=1,\ldots,D\},
\eea
is an {\em orthogonal product basis} (OPB). 

The OPBs can be divided into reducible and irreducible ones.
We say that the OPB $\cA$ is {\em reducible} if for some 
$i\in\{1,2,\ldots,n\}$ there exists a partition of 
$\{1,2,\ldots,D\}$ into two nonempty parts $J$ and $K$ such that 
$\ket{a_{j,i}}\perp\ket{a_{k,i}}$ for all $j\in J$ and $k\in K$. 
An OPB is {\em irreducible} if it is not reducible.

Assume that $\cA$ is reducible and let $i,J,K$ be as above. 
Denote by $\cH_{i,J}$ the subspace of $\cH_i$ spanned by 
the vectors $\ket{a_{j,i}}$ with $j\in J$, and define 
$\cH_{i,K}$ similarly. Denote by $\cH'$ the subsystem 
$\cH_1\ox\cdots\ox\cH_{i,J}\ox\cdots\ox\cH_n$ of $\cH$, 
and define $\cH''$ similarly (by replacing $J$ with $K$). 
Clearly, $\cA$ is the union of an OPB of $\cH'$ and 
an OPB of $\cH''$. This shows that the description of the 
OPBs of the system $\cH$ reduces to the case of irreducible 
OPBs (modulo the systems of lower dimension). 
The irreducible OPBs in the case of two qutrits and three qubits have been classified in \cite{fs09}. 

For a nonzero vector $\ket{x}\in\cH_i$, we denote by $[x]$ the 
one-dimensional subspace spanned by this vector. 
If $\{\ket{x_j}:j=1,\ldots,d_i\}$ is a basis of $\cH_i$, 
then we say that 
$$
F:=\{ [x_j]: j=1,2,\ldots,d_i \} 
$$
is a \emph{frame} of $\cH_i$. If moreover the $[x_j]$ are 
pairwise orthogonal, we say that $F$ is an 
\emph{orthogonal frame}. 

For the OPB $\cA$ given by \eqref{eq:bazis-A}, we set 
\bea \label{eq:cA_i}
\cA_i=\{[a_{s,i}]:s=1,\ldots,D\}, \quad i=1,\ldots,n.
\eea
For any 1-dimensional subspace $V\subseteq\cH_i$ we define its 
\emph{multiplicity} $\mu(V)$ to be the number of indices 
$s\in\{1,2,\ldots,D\}$ such that $[a_{s,i}]=V$. In particular, 
$\mu(V)=0$ if $[a_{s,i}]\ne V$ for all indices $s$.

\bpp \label{pp:OPB} 
Suppose that $d_1=2$ and set $\cH'=\cH_2\ox\cdots\ox\cH_n$. 
Let $\cA$ be the OPB given by \eqref{eq:bazis-A} and define 
the $\cA_i$ by \eqref{eq:cA_i}. For each $s\in\{1,2,\ldots,D\}$ 
set $\ket{a'_s}:=\ket{a_{s,2},a_{s,3},\ldots,a_{s,n}}$. 
Let $\cE=\{V_1,V_2,\ldots,V_m\}$ ($V_j$ distinct) be any maximal subset of $\cA_1$ not containing any orthogonal frame. Then the following hold.

(i) $\cA_1$ is the disjoint union of orthogonal frames 
$F_j=\{V_j,V_j^\perp\}$, $j=1,2,\ldots,m$, 
$\mu(V_j)=\mu(V_j^\perp)$ for each $j$, and the vectors 
$\ket{a'_s}$ for which $[a_{s,1}]\in\cE$ form an OPB of $\cH'$.

(ii) If $P_j=\{s:[a_{s,1}]=V_j\}$ and 
$Q_j=\{s:[a_{s,1}]=V_j^\perp\}$ then 
$\sum_{s\in P_j} [a'_s]=\sum_{s\in Q_j} [a'_s]$.

(iii) If all $d_i=2$ then $\sum_{i=1}^n \mu([a_{s,i}])\ge D-1$
for each $s\in\{1,2,\ldots,D\}$.
\epp
\bpf
(i) Let us first assume that $\cE$ is chosen so that 
$\mu(V_j)\ge\mu(V_j^\perp)$ for each $j$. 
The maximality property implies that for each 
$V\in\cA_1\setminus\cE$ we have $V^\perp\in\cE$.
Let $\mu'=\sum_{V\in\cE} \mu(V)$ and 
$\mu''=\sum_{V\in\cE} \mu(V^\perp)$. Note that 
$\mu'+\mu''=\sum_{V\in\cA_1} \mu(V)=D$  
and $\mu'-\mu''=\sum_{V\in\cE}(\mu(V)-\mu(V^\perp)\ge0$. 
If $[a_{s,1}],[a_{t,1}]\in\cE$, $s\ne t$, then 
$\braket{a_{s,1}}{a_{t,1}}\ne0$ and so 
$\ket{a'_s}\perp\ket{a'_t}$. It follows that 
$\sum [a'_s]$, taken over all indices $s$ for which 
$[a_{s,1}]\in\cE$, has dimension $\mu'$. 
Consequently, $\mu'\le\dim \cH'=D/2$. Hence, 
$2\mu'\le D=\mu'+\mu''$, i.e., $\mu'\le\mu''$. As also
$\mu''\le\mu'$, we have $\mu'=\mu''$. It follows that 
for each $V\in\cE$ we have $\mu(V)=\mu(V^\perp)$, and so 
$V^\perp\in\cA_1$. 
Consequently, all possible maximal subsets $\cE\subseteq\cA_1$ 
not containing any orthonormal frame satisfy the additional
assumption made at the beginning of the proof. 
We can now drop that assumption.

The first two assertions of (i) have been proved. 
The third follows from the fact shown above that 
$\ket{a'_s}\perp\ket{a'_t}$ provided that 
$[a_{s,1}],[a_{t,1}]\in\cE$ and $s\ne t$. 

(ii) From (i) we know that $\cH'$ is an orthogonal direct sum 
of the subspaces $X_j:=\sum_{s\in P_j} [a'_s]$, $j=1,2,\ldots,m$,
and also orthogonal direct sum of the subspaces 
$Y_j:=\sum_{s\in Q_j} [a'_s]$, $j=1,2,\ldots,m$. Since 
$\mu(V_k)=\mu(V_k^\perp)$ we can interchange the roles of
$V_k$ and $V_k^\perp$ for a single index $k$. 
Hence $Y_k$ is also the orthogonal complement of 
$\sum_{j\ne k} X_j$ and so $Y_k=X_k$. 

(iii) For each $t\ne s$ there is at least one $i$ such that 
$\braket{a_{s,i}}{a_{t,i}}=0$. Let $J_i$ be the set of indices 
$t$ such that $\braket{a_{s,i}}{a_{t,i}}=0$. As all $d_i=2$, 
$\braket{a_{s,i}}{a_{t,i}}=0$ is equivalent to 
$[a_{t,i}]=[a_{s,i}]^\perp$.
Since $\mu([a_{s,i}])=\mu([a_{s,i}]^\perp)$, we have 
$|J_i|=\mu([a_{s,i}])$. As $\cA$ is an OPB, we have 
$\cup_{i=1}^n J_i = \{1,2,\ldots,D\} \setminus \{s\}$. Hence 
$\sum_{i=1}^n \mu([a_{s,i}])=\sum_{i=1}^n |J_i|\ge D-1$.
\epf

It follows from this proposition that in the bipartite 
systems with $d_1=2$ we can construct all OPBs by the 
following method. We choose an orthogonal decomposition 
$\cH_2=X_1\oplus \cdots \oplus X_m$ and for each 
$j\in\{1,2,\ldots,m\}$ we choose a unit vector 
$v_j\in\cH_1$ and two arbitrary o.n. bases 
$\{ x_{j,1},x_{j,2},\ldots,x_{j,k_j}\}$ and 
$\{ y_{j,1},y_{j,2},\ldots,y_{j,k_j}\}$ of $X_j$. 
Then the $D$ product vectors 
\begin{eqnarray*}
&&
\ket{v_j,x_{j,1}},\ldots,\ket{v_j,x_{j,k_j}}, \\
&&
\ket{v_j^\perp,y_{j,1}},\ldots,\ket{v_j^\perp,y_{j,k_j}}, \\
&&
j=1,\ldots,m,
\end{eqnarray*}
form an OPB of $\cH$.

\bcr \label{cr:n=d_1=2}
Any OPB of the bipartite system $2\ox d_2$ is reducible. 
\ecr
\bpf
The assertion is obvious if $m=1$. If $m>1$ it follows from 
the observation that the first $2k_1$ product vectors in the above list (those with $j=1$) are orthogonal to all the remaining product vectors in the list.
\epf

This result also follows from \cite[Theorem 3]{fs09}, which says that any irreducible OPB is LOCC-indistinguishable, and the 
known fact that any $2\ox d_2$ OPB is LOCC-distinguishable  
(see the end of \cite{bdm99}).
It does not extend to other bipartite systems. 
For instance, in the case $d_1=d_2=3$ 
there exist irreducible OPBs \cite[Fig 1]{fs09}.

In Proposition \ref{pp:OPB} (i), $\cA_2$ is not necessarily 
a disjoint union of orthogonal frames. For instance, 
let $d_2=3$ and consider the OPBs of the form 
$\ket{0,0}$, $\ket{0,1}$, $\ket{0,2}$, 
$\ket{1,0}$, $\ket{1,x}$, $\ket{1,y}$ with $[x]\ne[1],[2]$.

\section{OPBs of multiqubit systems} 
\label{sec:combinatorial}

In this section we reduce the classification problem of OPBs 
in multiqubit systems to a purely combinatorial problem. 
Thus we set $d_1=d_2=\cdots=d_n=2$ and so $D=2^n$. 

Given a unit product vector $\ket{x_1,x_2,\ldots,x_n}$, we shall 
always assume (as we may) that the vectors $\ket{x_j}$ are unit 
vectors. 
For convenience, in this section we shall not distinguish two unit vectors in $\cH_j$ which differ only by a phase factor, i.e., we consider these vectors as points of the complex 
projective line $\bP(\cH_j)$ associated to $\cH_j$.
If $\ket{x}\in\cH_j$ is a unit vector, then by using this 
convention, we can say that there exists a unique unit vector $\ket{x}^\perp\in\cH_j$ which is orthogonal to $\ket{x}$. 
We refer to $\ket{x}^\perp$ as the \emph{perpendicular} 
of $\ket{x}$.

Let $\Un(\cH_i)$ be the unitary group of the 2-dimensional Hilbert space $\cH_i$. For $U_i\in\Un(\cH_i)$, $i=1,2,\ldots,n$, let $U=(U_1,U_2,\ldots,U_n)$ be the corresponding element 
of the direct product of the groups $\Un(\cH_i)$. 
Then $U$ acts on $\cH$ as the {\em local unitary operator} 
$U_1\ox U_2\ox \cdots \ox U_n$. Thus if 
$\ket{x}=\ket{x_1,x_2,\ldots,x_n}$, we have  
$U\ket{x}=U_1\ket{x_1}\ox U_2\ket{x_2}\ox \cdots 
\ox U_n\ket{x_n}$. Since we have fixed o.n. bases in all 
$\cH_i$, the symmetric group ${\rm Sym}_n$ acts on $\cH$ 
by permuting the tensor factors $\cH_i$. Thus 
$\pi\ket{x}=\ket{x_{\pi^{-1}(1)},x_{\pi^{-1}(2)},\ldots,
x_{\pi^{-1}(n)}}$ for $\pi\in{\rm Sym}_n$. 

\bd \label{def:OPB-equiv}
We say that two OPBs (or two families of OPBs) $\cA$ and $\cB$ are {\em equivalent} if there exist a local unitary operator $U$ and a permutation $\pi$ such that $\cB=\pi U\cA$.
\ed

(If $\cA$ is a family of OPBs then $\pi U\cA$ denotes the family 
obtained by applying the operator $\pi U$ to each member of
the family $\cA$.)

We denote by $\cH_\times$ the product of the Hilbert spaces 
$\cH_i$, $\cH_\times=\cH_1\times\cH_2\times\cdots\times\cH_n$. 
Further, we denote by $\cH_\times^r$ the product of the 
$r$ copies of the space $\cH_\times$. 

We can represent an OPB, say 
$\cA=\{\ket{a_{i,1},a_{i,2},\ldots,a_{i,n}}\}$, 
by the corresponding point of the space $\cH_\times^{2^n}$, 
i.e., the $2^n\times n$ matrix $A$ with rows 
$[~ \ket{a_{i,1}}~\ket{a_{i,2}}~\cdots~\ket{a_{i,n}} ~]$. 
We are interested in the equivalence classes of OPBs for the 
equivalence defined above. In particular, this means that we can permute the columns of $A$. Since an OPB is just a set (not an ordered set), we can also permute the rows of $A$. Let us give an example of a local unitary operation.
The product vectors $\ket{0,0},\ket{0,1},\ket{1,+},\ket{1,-}$  form an OPB of a 2-qubit system where 
$\ket{\pm}={1\over\sqrt2}(\ket{0}\pm\ket{1})$. After applying a Hadamard gate on the second qubit, one obtains $\ket{0,+},\ket{0,-},\ket{1,0},\ket{1,1}$, which is also an OPB. In this sense, the above two OPBs are locally unitarily equivalent but not obtained by just permuting the rows of $A$.

We will show that the OPBs occur, up to equivalence, in several infinite families which will be specified by $2^n\times n$ matrices $M$ whose entries are unit vectors considered as variables. Let us say that $\ket{v}$ is a {\em vector variable} if it runs through all unit vectors in one of the spaces $\cH_j$. 
If $\ket{v}$ is a vector variable on $\cH_j$ then the same
is true for its perpendicular. We say that a finite collection 
of pairwise distinct vector variables on $\cH_j$ is 
{\em independent} if it does not contain a pair consisting of a 
vector variable and its perpendicular.

\bd
We define $\cO=\cO(n)$ formally to be the set of $2^n\times n$ 
matrices $M=[M_{i,k}]$ having the following three properties:

(i) Each entry of the $j$th column of $M$ is either a (unit) vector variable, say $\ket{a}\in\cH_j$, or $\ket{a}^\perp$. 
To simplify notation, we shall write just $a$ and $a^\perp$, 
respectively. Note that $(a^\perp)^\perp=a$.

(ii) If a vector variable, say $a$, occurs in a column of $M$ 
then neither $a$ nor $a^\perp$ occur in any other column.

(iii) Any two distinct rows of $M$, say rows $i$ and $j$, are 
\emph{orthogonal} to each other in the sense that
$M_{i,k}=M_{j,k}^\perp$ for some $k$.
\ed

For a given $M\in\cO(n)$, we denote by $\mu(a)$ the number of 
occurencies of the vector variable $a$ in the matrix $M$. We 
refer to $\mu(a)$ as the {\em multiplicity} of $a$. Thus if 
a vector variable $a$ does not occur in $M$, then $\mu(a)=0$. 
We shall prove that $\mu(a)=\mu(a^\perp)$.

\bl \label{Posledica}
A vector variable $a$ and its perpendicular $a^\perp$ occur in $M\in\cO(n)$ the same number of times, i.e., we have 
$\mu(a)=\mu(a^\perp)$. 
\el
\bpf
Let $\{a_1,a_2,\ldots,a_m\}$ be a maximal set of independent 
variables which occur in the first column of $M$. Without any 
loss of generality, we may assume that this set is chosen so 
that $\mu(a_i)\ge\mu(a_i^\perp)$ for all $i$. Set 
$\mu':=\sum_{i=1}^m \mu(a_i)$ and 
$\mu'':=\sum_{i=1}^m \mu(a_i^\perp)$ and note that 
$\mu'+\mu''=2^n$ and $\mu'\ge\mu''$.
Denote by $N$ the submatrix of $M$ obtained by first deleting all rows whose first element is one of the variables $a_i^\perp$ 
and then deleting the first column. Thus $N$ has $\mu'$ rows and 
$n-1$ columns. Moreover the rows of $N$ are mutually orthogonal.
This implies that $\mu'\le 2^{n-1}$. Hence we must have 
$\mu'=\mu''=2^{n-1}$ and $\mu(a_i)=\mu(a_i^\perp)$ for each $i$.
\epf

To define equivalence of matrices $M\in\cO(n)$ we need to 
rename some vector variables. This may be confusing, so we 
first describe a {\em simple renaming}. Let $a$ be a vector 
variable which occurs in $M$. Recall that $a$ and $a^\perp$ 
occur only in a single column of $M$, say column $j$, and that 
$\mu(a)=\mu(a^\perp)$. We choose a new vector variable 
$x$ on $\cH_j$. Finally, we replace simultaneously each occurrence of $a$ with $x$ and each occurrence of $a^\perp$ 
with $x^\perp$. A general {\em renaming} is just a composition of finitely many simple renamings. In particular it may be 
a trivial renaming, which means that we do not make any 
changes in $M$.

\bd
\label{def:equivalent}
We say that two matrices $M,N\in\cO(n)$ are {\em equivalent} 
if $N$ can be obtained from $M$ by permuting rows and 
columns and renaming of the vector variables. We refer to 
row permutations, column permutations and renamings as 
{\em equivalence operations}.
\ed

We denote by $[A]$ the equivalence class in $\cO(n)$ containing the matrix $A\in\cO(n)$.

Let us give an example. For instance the matrices $A,X\in\cO(2)$ given by
\bea 
A
=\left[ \begin{array}{cc}
a & b \\
a & b^\perp \\
a^\perp & c \\ 
a^\perp & c^\perp 
\end{array} \right], \qquad
X
=\left[ \begin{array}{cc}
x & z \\
y & z^\perp \\
y^\perp & z^\perp \\ 
x^\perp & z
\end{array} \right]
\eea 
are equivalent, i.e., we have $[A]=[B]$. The reason is that we can transform $A$ to $X$ by the three transformations
\bea 
\left[ \begin{array}{cc}
a & b \\
a & b^\perp \\
a^\perp & c \\ 
a^\perp & c^\perp 
\end{array} \right] \ra
\left[ \begin{array}{cc}
b & a \\
b^\perp & a \\
c & a^\perp \\ 
c^\perp & a^\perp 
\end{array} \right] \ra
\left[ \begin{array}{cc}
b^\perp & a \\
c & a^\perp \\ 
c^\perp & a^\perp \\ 
b & a
\end{array} \right] \ra
\left[ \begin{array}{cc}
x & z \\
y & z^\perp \\
y^\perp & z^\perp \\ 
x^\perp & z
\end{array} \right].
\eea 
The first transformation is a column permutation, 
the second a row permutation and the third is the renaming 
of $a,b,c$ to $z,x^\perp,y$ respectively.

We say that $M\in\cO(n)$ is {\em irreducible} if each column of $M$ contains at least two independent vector variables. 
We say that $M$ is {\em reducible} if it is not irreducible.

Let $s_j$ be a vector variable on $\cH_j$, $j=1,2,\ldots,n$. 
For $i=1,2,\ldots,2^n$ we write the integer $i-1$ in base 2 
as $i-1=\sum_{j=1}^n d_{i,j} 2^{j-1}$, where $d_{i,j}\in\{0,1\}$ are the binary digits. We define the {\em standard matrix} 
$S:=[s_{i,j}]\in\cO(n)$ by setting $s_{i,j}=s_j$ if 
$d(i,j)=0$ and $s_{i,j}=s_j^\perp$ if $d(i,j)=1$. 
We refer to the equivalence class $[S]$ as the {\em standard class}. It is easy to see that all matrices $A\in\cO(n)$, which have the property that $\mu(x)=2^{n-1}$ for each entry $x$ of 
$A$, belong to the standard class. All matrices in this class are obviously reducible.

Let $M\in\cO$ and let $a$ and $b$ be independent vector variables which occur in the same column of $M$. Then the
matrix obtained from $M$ by setting $b=a$ and $b^\perp=a^\perp$ 
everywhere in $M$ or by setting $b=a^\perp$ and $b^\perp=a$ 
everywhere in $M$ also belongs to $\cO$. 
If $N$ can be obtained from $M$ by this procedure we shall 
write $N<M$. 
Note that, according to this definition, $M<N$ and $N<P$ do not imply that $M<P$, i.e., the binary relation ``$<$'' is not transitive.

\bd
\label{def:maximal}
We say that $M\in\cO$ is {\em maximal} if there is no 
$N\in\cO$ such that $M<N$.
\ed

Since each matrix $M\in\cO$ arises from some maximal 
matrix $N\in\cO$ by identification of some vector variables,
the construction of OPBs of the $n$-qubit system reduces to the enumeration of the equivalence classes of maximal matrices $M\in\cO$. Hence, in order to construct the OPBs of $\cH$, it 
suffices to classify (up to equivalence) the maximal matrices 
$M\in\cO$.

Let $A=[a_{i,j}]$ be an $s\times t$ matrix whose entries are vector variables. We say that the rows of $A$ are 
{\em independent} if we can assign unit vectors to these variables so that the product vectors
$\ket{a_{i,1},a_{i,2},\ldots,a_{i,t}}$, $i=1,2,\ldots,s$, 
are linearly independent.

\bl \label{Ortog}
If $A=[a_{i,j}]$ is an $s\times n$ matrix of vector variables 
whose rows are independent, then $s\le2^n$. 
\el
\bpf
This follows from the fact that $\dim \cH=2^n$.
\epf

\bex
\label{eg:2qubit}
{\rm
Let $A\in\cO(2)$ and let $[a ~ b]$ be its first row. By 
Corollary \ref{cr:n=d_1=2}, $A$ is reducible. Hence, one of the 
columns contains only one independent variable. By interchanging the columns if necessary, we may assume that the first column 
contains only one independent variable. Since 
$\mu(a)=\mu(a^\perp)$, by permuting the rows, we may assume 
that the first column is $[a~a~a^\perp~a^\perp]^T$. 
Since the first row is orthogonal to the second, we deduce that
the second row must be $[a~b^\perp]$. There are now two choices 
for the remaining two entries of $A$. They lead to the two 
matrices $M$ and $N$ shown below. Thus, in the case of two qubits there are only two equivalence classes in $\cO$. Their representatives are
\bea \label{mat:2-q}
M
=\left[ \begin{array}{cc}
a & b \\
a & b^\perp \\
a^\perp & b \\ 
a^\perp & b^\perp 
\end{array} \right], \quad
N
=\left[ \begin{array}{cc}
a & b \\
a & b^\perp \\
a^\perp & c \\ 
a^\perp & c^\perp 
\end{array} \right].
\eea 
Both matrices are reducible, $M$ belongs to the standard class  and $N$ is maximal. Since $M$ is 
obtained from $N$ by setting $c=b$, we have $M<N$.
\qed
}
\eex

We shall associate to $M\in\cO(n)$ a family $\cF_M$ of OPBs 
of $\cH$. To do that, we assign to all vector variables in $M$ 
unit vectors. It is understood that to a pair $x,x^\perp$ which 
occur in column $j$, we assign a pair of orthogonal unit vectors in $\cH_j$. Denote by $\ket{\alpha_i}\in\cH$ the product vector which is the tensor product of the unit vectors assigned to the entries of the row $i$ of $A$. Then 
$\{\ket{\alpha_i}:i=1,2,\ldots,2^n\}$ is an OPB of $\cH$. The family $\cF_M$ consists of all OPBs that arise from $M$ in this way. 

For instance, if $N$ is the matrix displayed above in 
\eqref{mat:2-q}, then $\cF_N$ consists of all OPBs 
$\{ \ket{x}\ox\ket{y}, \ket{x}\ox\ket{y}^\perp, 
\ket{x}^\perp\ox\ket{z}, \ket{x}^\perp\ox\ket{z}^\perp \}$,
where $\ket{x}\in\cH_1$ and $\ket{y},\ket{z}\in\cH_2$ are arbitrary unit vectors.

Note that two matrices $M,N\in\cO$ are equivalent if and 
only if the corresponding families $\cF_M$ and $\cF_N$ are equivalent under local unitary transformations and qubit permutations. If $M$ is irreducible then, in the generic case, the members of $\cF_M$ are irreducible OPBs. 

If $S\in\cO(n)$ is a standard matrix, then we say that the family $\cF_S$ is the {\em standard family}. This family consists of all OPBs which are equivalent to the standard basis of $\cH$. 
Given any matrix $A\in\cO(n)$, there is a finite chain 
$S=A_0 < A_1 < \cdots < A_m = A$, $m\ge0$, 
which begins with a standard matrix $S$ (on some variables) and reaches $A$. Consequently, we have $\cF_S\subseteq\cF_A$.

One can use the OPBs in $\cF_M$ to derive some properties 
of the matrix $M$. We illustrate this by a simple lemma.

\bl \label{le:1-row}
If the matrices $M,N\in\cO(n)$ have all rows equal except possibly the first, then $M=N$.
\el
\bpf
Let us assign to all vector variables that occur in $M$ and $N$ 
different unit vectors. The corresponding OPBs will consist 
of the same vectors except possibly one of them. As they are 
orthonormal bases, these two OPBs must be the same. Thus the 
first rows of $M$ and $N$ give the same product vector. 
As we have assigned different unit vectors to different 
vector variables, we infer that the first rows of $M$ and $N$ 
must be the same.
\epf

It is easy to see that any OPB of $\cH$ belongs to some family $\cF_M$, $M\in\cO$. Indeed, let 
$\cA:=\{\ket{a_s}=\ket{a_{s,1},\ldots,a_{s,n}}:s=1,\ldots,2^n\}$
be an OPB, and let $A$ be the $2^n\times n$ matrix whose entries 
are the unit vectors $A[s,j]=\ket{a_{s,j}}$. Note that several 
entries in a column $j$ of $A$ may be equal to some $\ket{a_{s,j}}$. We replace all of them by a single vector variable $v$, and likewise replace all the entries equal to 
$\ket{a_{s,j}}^\perp$ with $v^\perp$. The resulting matrix 
$M$ has vector variables as its entries and belongs to 
$\cO$. Moreover, we have $\cA\in\cF_M$.

\section{Weak equivalence in $\cO$}
\label{sec:weakequiv}

In the previous section we have reduced the problem of constructing the multiqubit OPBs to a combinatorial problem. In this section, we investigate the latter problem. We begin by introducing two more definitions. 
We say that a collection of rows of $M\in\cO=\cO(n)$ is 
{\em j-constant} if all entries of the column $j$ contained in these rows are equal to each other. For a subset 
$J\subseteq\{1,\ldots,n\}$ we say that a collection of rows of $M$ is \emph{J-compatible} if these rows are $j$-constant for all $j\notin J$.

\bpp
\label{pp:sused}
Let $M\in\cO$ be a maximal matrix, 
$J\subset\{1,2,\ldots,n\}$ a subset of cardinality $k>0$,
and $I\subset\{1,2,\ldots,2^n\}$ a subset 
of cardinality $2^k$ such that 
the rows of $M$ with indices in $I$ are $J$-compatible. 
Denote by $B$ the submatrix of $M$ contained in the intersection 
of rows $I$ and columns $J$. Then

(i) $B\in\cO(k)$;

(ii) for $i\notin I$, the portion of row $i$ contained in 
columns $J$ is not orthogonal to all the rows of $B$;

(iii) any vector variable that occurs in $B$ does not occur 
in $M$ outside of $B$;

(iv) the submatrix $B$ is maximal in $\cO(k)$;

(v) the matrix $M'$ obtained from $M$ by permuting the  
columns of $B$ also belongs to $\cO$ and it is maximal.
\epp
\bpf
(i) Since the rows $I$ of $M$ are $J$-compatible, we deduce that all rows of $B$ must be mutually orthogonal and so $B\in\cO(k)$.

(ii) This follows from Lemma \ref{Ortog}.

(iii) Assume that a vector variable, say $a$, occurs in $B$ and 
also outside $B$. Let $N$ be the matrix obtained from $M$ by replacing each ocurrence of $a$ and $a^\perp$ inside $B$ by a 
new vector variable and its perpendicular, respectively. 

We claim that $N\in\cO$. We have to verify that $N$ satisfies 
the conditions (i-iii) of $\cO$. The conditions (i) and (ii) 
obviously hold. In order to verify the condition (iii) it 
suffices to show that if $m\notin I$ then the row $m$ of $M$ 
is orthogonal to all rows of $N$ in $I$. By part (ii) there 
exists $i\in I$ such that the portion of row $m$ in $J$ is not 
orthogonal to the row $i$ of $B$. Hence, the portion of row 
$m$ outside $J$ must be orthogonal to the corresponding portion 
of the row $i$ of $M$. Since the rows $I$ of $M$ are 
$J$-compatible, we conclude that row $m$ is orthogonal to 
all rows of $N$ in $I$. Thus our claim is proved. 
Obviously we have $M<N$, which contradicts the hypothesis that 
$M$ is maximal.

(iv) follows from (iii) and the maximality of $A$.

(v) follows immediately from the previous assertions.
\epf

We refer to the operation $M\to M'$ described in the above 
proposition as a {\em switching operation}. Let us give an 
example.

Assume that a maximal matrix $M\in\cO$ has a $4\times2$ 
submatrix $X$ 
contained in columns $k$ and $l$ such that the four rows 
containing this submatrix are  $\{k,l\}$-compatible. 
Then $X\in\cO(2)$ and, by using a switching operation, 
we can replace the submatrix $X$ by the matrix $Y\in\cO(2)$ 

\bea
X=\left[ \begin{array}{cc}
a&x\\
a&x^\perp\\
a^\perp & y \\
a^\perp & y^\perp \\
\end{array} \right] \to
Y=\left[ \begin{array}{cc}
x&a\\
x^\perp&a\\
y&a^\perp\\
y^\perp&a^\perp\\
\end{array} \right] 
\eea
to obtain another maximal matrix $M'\in\cO$.

\bd
\label{def:w-equivalent}
We say that two maximal matrices $M,N\in\cO(n)$ are 
{\em weakly equivalent} if there is a finite sequence 
$M=M_0,M_1,\ldots,M_k=N$ in $\cO$ such that each arrow 
$M_{i-1}\to M_i$ is an equivalence or switching operation.
We shall refer to the equivalence classes of the weak 
equivalence relation as {\em switching classes}. 
\ed

Each switching class in $\cO$ consists of maximal matrices 
and it is a disjoint union of finitely many previously defined equivalence classes. The construction of matrices in $\cO(n)$ can be carried out in two steps: first find the 
representatives of the switching classes, and then find the representatives of the equivalence classes contained in each switching class. 

By the above definition, two equivalent maximal matrices are 
also weakly equivalent. The converse is false. For example the reducible maximal matrices of three qubits form a single 
switching class which is the union of two equivalence 
classes (see Lemma  \ref{le:3qubits} below). 
It follows from Proposition \ref{pp:sused} (iii) that a 
switching class cannot contain a reducible and a irreducible 
matrix.

When displaying matrices $M$ we shall use some shorthand 
notation in order to diminish the number of rows. 
It is also convenient to specify one of the vector variables 
in some column to be the standard basis vector $\ket{0}$.
For instance, if $a$ occurs in column $j$ then we can replace in that column each $a$ with $0$ and each $a^\perp$ with $1$. This reduces the number of vector variables by one. We say that the column $j$ of the resulting matrix is \emph{normalized}. 
Note that this normalization is not unique.

We often simplify a maximal matrix $M$ by using the symbol *. 
Assume that a vector variable, say $a$, occurs in row $i$ and 
column $k$ of $M$. Then $a^\perp$ also occurs in column $k$ and, say, row $j$. Assume also that the rows $i$ and $j$ are 
$\{k\}$-compatible. Since $M$ is maximal, we must have 
$\mu(a)=1$ by Proposition \ref{pp:sused} (iii). Under these assumptions we can replace 
$a$ in row $i$  with $*$ and delete row $j$. 
We can recover (up to ordering of the rows and naming of the vector variables) the original $M$ from this simplified matrix
by reversing this procedure. 
For instance, if $n=4$ then the symbolic row $[~*~b~c~d~]$ is a shorthand for the pair of rows
$$
\left[ \begin{array}{cccc}
a & b & c & d \\
a^\perp & b & c & d 
\end{array} \right].
$$
For a concrete example see \eqref{eq:3qubit=8row} where 
we simplified the matrix $M_{\rm nor}$ to get a matrix 
with 5 rows only. 

We may apply this simplification several times one after the other. For instance, when $n=4$ the two rows 
\bea \label{eq:2rows}
\left[ \begin{array}{cccc}
* & a & b & c \\
* & a & b & c^\perp 
\end{array} \right].
\eea
stand for the following four rows
$$
\left[ \begin{array}{cccc}
d & a & b & c \\
d^\perp & a & b & c \\
e & a & b & c^\perp \\
e^\perp & a & b & c^\perp 
\end{array} \right],
$$
where $d$ and $e$ are distinct two new independent vector variables.

Let us give two small examples. 
\bex \label{ex:2i3kubita}
{\rm
First, in the case of two qubits there is only one 
maximal matrix $N\in\cO$ up to equivalence. 
This is the matrix shown in \eqref{mat:2-q}. 
Its normalized version is 
$$
\left[ \begin{array}{cc}
0 & 0 \\
1 & 0 \\ 
b & 1 \\
b^\perp & 1 
\end{array} \right],
$$
where we have specified that $\ket{a}=\ket{0}$ and 
$\ket{c}=\ket{0}$. Note that $M$ is reducible since 
its second column contains only one independent 
vector variable.

Second, according to \cite{fs09} there is a unique family of irreducible three-qubit OPBs. In our notation, this family 
is given by the following matrix $M$
\bea
\label{eq:3qubit=8row}
M=\left[ \begin{array}{ccc}
u & v & w \\
a & v^\perp & w \\
a^\perp & v^\perp & w \\
u & b & w^\perp \\
u & b^\perp & w^\perp \\
u^\perp & v & c \\
u^\perp & v & c^\perp \\
u^\perp & v^\perp & w^\perp 
\end{array} \right], \qquad
M_{\rm nor}=\left[ \begin{array}{ccc}
0 & 0 & 0 \\
a & 1 & 0 \\
a^\perp & 1 & 0 \\
0 & b & 1 \\
0 & b^\perp & 1 \\
1 & 0 & c \\
1 & 0 & c^\perp \\
1 & 1 & 1 
\end{array} \right], \qquad
\left[ \begin{array}{ccc}
0 & 0 & 0 \\
* & 1 & 0 \\
0 & * & 1 \\
1 & 0 & * \\
1 & 1 & 1 
\end{array} \right].
\eea
The matrix $M_{\rm nor}$ is the normalization of $M$.
\qed
}
\eex

Apart from the asterisks and various vector variables $a$ and their perpendiculars $a^\perp$, all other entries of $M$ 
(when displayed) are the standard basis 
vectors $\ket{0}$ and $\ket{1}$ of the $\cH_i$. 
(These standard basis vectors are introduced by the 
normalization process mentioned above.)

Let us now consider the case $n=3$.

\bl \label{le:3qubits}
In $\cO(3)$ there are only two switching classes of maximal 
matrices: one of them consists of irreducible and the other
of reducible matrices. The former is a single equivalence 
classcas with representative \eqref{eq:3qubit=8row} while the latter splits into two equivalence classes with
representatives 
\bea
\label{eq:3qubit}
\left[\begin{array}{ccc}
a & b & c \\
a & b & c^\perp \\
a & b^\perp & d \\
a & b^\perp & d^\perp \\
a^\perp & e & f \\
a^\perp & e & f^\perp \\
a^\perp & e^\perp & g \\
a^\perp & e^\perp & g^\perp \\
\end{array}
\right],
~~~
\left[\begin{array}{ccc}
a & b & c \\
a & b & c^\perp \\
a & b^\perp & d \\
a & b^\perp & d^\perp \\
a^\perp & f & e \\
a^\perp & f^\perp & e \\
a^\perp & g & e^\perp \\
a^\perp & g^\perp & e^\perp \\
\end{array}
\right].
\eea 
\el
\bpf
In view of the Example \ref{ex:2i3kubita}, it suffices to 
consider the case when $M\in\cO(3)$ is reducible. We may assume 
that the first column of $M$ contains a vector variable $a$ 
with multiplicity $\mu(a)=4$. After permuting the rows, we may assume that

\bea
\label{eq:3kubita}
M = \left[\begin{array}{c|c}
a &  \\
a & M' \\
a & \\
a & \\
\hline
a^\perp & \\
a^\perp & M'' \\
a^\perp &  \\
a^\perp &  \\
\end{array}
\right].
\eea 
Since $M$ is maximal, Proposition \ref{pp:sused} (i) implies 
that $M',M''\in\cO(2)$. Moreover these two submatrices have no vector variable in common. By using  Example \ref{ex:2i3kubita}, it is now easy to verify that there are only two possibilities (up to equivalence) as stated in the lemma.
\epf

Let us introduce additional notation and invariants which will be used when testing whether two matrices in $\cO(n)$ are 
equivalent.

Given a matrix $M\in\cO(n)$, we denote by $\nu_i$ the number of independent vector variables which occur in column $i$ of $M$. Assume that these variables are $a_j$, $j=1,2,\ldots,\nu_i$, 
and set $\mu_{i,j}=\mu(a_j)$. 
We assume that the indexing is chosen so that  $\mu_{i,j}\ge \mu_{i,k}$ for $j<k$, and we set $\mu_i=\mu_{i,1}$.
The numbers $\mu_{i,1},\mu_{i,2},\ldots,\mu_{i,\nu_i}$
form a partition $\pi_i$ of the integer $2^{n-1}$. We shall order these partitions in the decreasing lexicographic order.

By permuting the columns of $M$, we may assume that 
$\pi_1\ge\pi_2\ge\cdots\ge\pi_n$.
In particular $\mu_i\ge \mu_j$ for $i<j$.

Another important invariant of matrices $M\in\cO(n)$ is the total number $\nu_M=\sum \nu_i$ of independent vector variables 
that occur in $M$. The dimension of the family $\cF_M$ is 
equal to $2\nu_M$ because each of the vector variables makes 
the contribution of 2 to this dimension.

By identifying two independent varables in a single column of 
one of the three maximal matrices \eqref{eq:3qubit=8row} and  \eqref{eq:3qubit}, and by repeating this procedure as far as possible we obatin a bunch of matrices in $\cO(3)$. By selecting 
a maximal subset of pairwise nonequivalent matrices in this 
bunch, we obtain the following corollary. The Hasse diagram of the equivalence classes of $\cO(3)$, for the partial order induced by the relation ``$<$'', is shown on Fig. 1.

\setlength{\unitlength}{1 mm}
\begin{figure}
\setlength{\unitlength}{1 mm}
\begin{center}
\begin{picture}(50,50)

\drawline(20,30)(30,40)
\drawline(20,30)(20,0)
\drawline(30,40)(40,30)
\drawline(30,40)(30,20)
\drawline(40,30)(40,20)
\drawline(40,20)(20,10)
\drawline(40,20)(20,0)
\drawline(30,20)(10,10)
\drawline(30,30)(10,10)
\drawline(20,0)(10,10)
\drawline(-20,20)(10,10)
\drawline(-20,20)(10,40)
\drawline(10,20)(10,10)
\drawline(10,20)(20,30)
\drawline(-10,20)(20,10)
\drawline(-10,20)(-20,30)
\drawline(-10,20)(20,30)
\drawline(10,40)(20,30)
\drawline(-5,30)(20,20)

\put(9,39){$\circ$}
\put(29,39){$\circ$}
\put(-21,29){$\circ$}
\put(-6,29){$\circ$}
\put(19,29){$\circ$}
\put(29,29){$\circ$}
\put(39,29){$\circ$}

\put(-21,19){$\circ$}
\put(-11,19){$\circ$}
\put(9,19){$\circ$}
\put(19,19){$\circ$}
\put(29,19){$\circ$}
\put(39,19){$\circ$}

\put(9,9){$\circ$}
\put(19,9){$\circ$}
\put(29,9){$\circ$}

\put(19,-1){$\circ$}

\put(41,28){\scriptsize$7$}
\put(41,20){\scriptsize$8$}
\put(30,41){\scriptsize$1$}
\put(31,28){\scriptsize$5$}
\put(31,20){\scriptsize$13$}
\put(31,8){\scriptsize$16$}
\put(21,28){\scriptsize$4$}
\put(21,18){\scriptsize$11$}
\put(21,7){\scriptsize$14$}
\put(21,-2){\scriptsize$17$}
\put(11,41){\scriptsize$2$}
\put(5,18){\scriptsize$12$}
\put(6,6){\scriptsize$15$}
\put(-7,31){\scriptsize$3$}
\put(-15,19){\scriptsize$10$}
\put(-23,29){\scriptsize$6$}
\put(-23,19){\scriptsize$9$}

\put(60,40){$\n=7$}
\put(60,30){$\n=6$}
\put(60,20){$\n=5$}
\put(60,10){$\n=4$}
\put(60,0){$\n=3$}

\end{picture}
\end{center}
\caption{Hasse diagram of the partially ordered set of equivalence classes in $\cO(3)$. The partial order is induced by the relation ``$<$''.}
\end{figure}
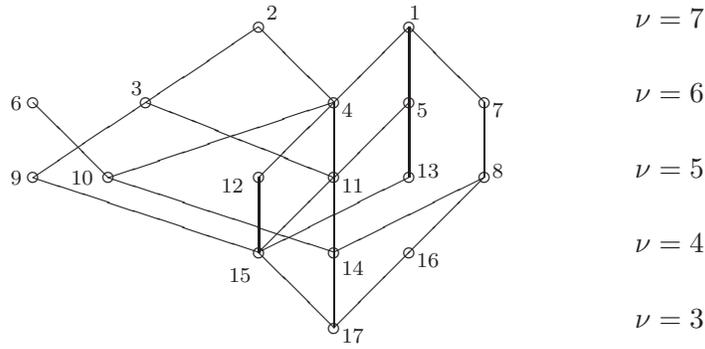

\begin{corollary} \label{cor:eq-cl3qub}
There are 17 equivalence classes in $\cO(3)$. Their representatives $M$ are listed below together with associated 
partitions $\pi_1,\pi_2,\pi_3$ and the parameter $\nu_M$.
(All columns but one are normalized. On Fig. 1, the equivalence classes are numbered according to their position in the list.)

\newpage 

\begin{eqnarray*}
\left[ \begin{array}{ccc}
0 & 0 & * \\
0 & 1 & * \\
1 & y & * \\
1 & y^\perp & * 
\end{array} \right] 
\begin{array}{l}4 \\ 2^2 \\ 1^4 \\
\nu=7 \end{array} \ 
\left[ \begin{array}{ccc}
0 & 0 & * \\
0 & 1 & * \\
1 & * & 0 \\
1 & * & 1
\end{array} \right]
\begin{array}{l}4 \\ 2,1^2 \\ 2,1^2 \\
\nu=7 \end{array} \ 
\left[ \begin{array}{ccc}
0 & 0 & * \\
0 & 1 & * \\
1 & * & 0 \\
1 & 0 & 1 \\
1 & 1 & 1
\end{array} \right]
\begin{array}{l}4 \\ 3,1 \\ 2,1^2 \\
\nu=6 \end{array} \\
\left[ \begin{array}{ccc}
0 & y & * \\
0 & y^\perp & * \\
1 & 0 & 0 \\
1 & 0 & 1 \\ 
1 & 1 & 0 \\ 
1 & 1 & 1
\end{array} \right]
\begin{array}{l}4 \\ 2^2 \\ 2,1^2 \\
\nu=6 \end{array} \
\left[ \begin{array}{ccc}
0 & 0 & 0 \\
0 & 0 & 1 \\
0 & 1 & * \\
1 & y & 0 \\
1 & y & 1 \\
1 & y^\perp & * 
\end{array} \right]
\begin{array}{l}4 \\ 2^2 \\ 2,1^2 \\
\nu=6 \end{array} \ 
\left[ \begin{array}{ccc}
0 & 0 & 0 \\
* & 1 & 0 \\
0 & * & 1 \\
1 & 0 & * \\
1 & 1 & 1 
\end{array} \right]
\begin{array}{l} 3,1 \\ 3,1 \\ 3,1 \\
\nu=6 \end{array} \\
\left[ \begin{array}{ccc}
0 & 0 & * \\
0 & 1 & * \\
1 & 0 & * \\
1 & 1 & * 
\end{array} \right]
\begin{array}{l}4 \\ 4 \\ 1^4 \\
\nu=6 \end{array} \ 
\left[ \begin{array}{ccc}
0 & 0 & 0 \\
0 & 0 & 1 \\
0 & 1 & * \\
1 & 0 & * \\
1 & 1 & 0 \\
1 & 1 & 1 
\end{array} \right]
\begin{array}{l} 4 \\ 4 \\ 2,1^2 \\
\nu=5 \end{array} \
\left[ \begin{array}{ccc}
0 & 0 & * \\
0 & 1 & * \\
1 & 0 & 0 \\
1 & 0 & 1 \\ 
1 & 1 & 0 \\ 
1 & 1 & 1
\end{array} \right]
\begin{array}{l} 4 \\ 4 \\ 2,1^2 \\
\nu=5 \end{array} \\
\left[ \begin{array}{ccc}
0 & 0 & 0 \\
0 & 1 & 0 \\
0 & * & 1 \\
1 & 0 & * \\ 
1 & 1 & 0 \\ 
1 & 1 & 1
\end{array} \right]
\begin{array}{l} 4 \\ 3,1 \\ 3,1 \\
\nu=5 \end{array} \ 
\left[ \begin{array}{ccc}
0 & 0 & 0 \\
0 & 0 & 1 \\
0 & 1 & 0 \\
0 & 1 & 1 \\
1 & * & z \\ 
1 & 0 & z^\perp \\ 
1 & 1 & z^\perp
\end{array} \right]
\begin{array}{l} 4 \\ 3,1 \\ 2^2 \\
\nu=5 \end{array} \ 
\left[ \begin{array}{ccc}
0 & 0 & 0 \\
0 & 0 & 1 \\
0 & 1 & 0 \\
0 & 1 & 1 \\
1 & y & z \\ 
1 & y & z^\perp \\ 
1 & y^\perp & z \\
1 & y^\perp & z^\perp 
\end{array} \right]
\begin{array}{l} 4 \\ 2^2 \\ 2^2 \\
\nu=5 \end{array} \\
\end{eqnarray*}

\begin{eqnarray*}
\left[ \begin{array}{ccc}
0 & 0 & 0 \\
0 & 0 & 1 \\
0 & 1 & z \\
0 & 1 & z^\perp \\
1 & y & 0 \\ 
1 & y & 1 \\ 
1 & y^\perp & z \\
1 & y^\perp & z^\perp 
\end{array} \right]
\begin{array}{l} 4 \\ 2^2 \\ 2^2 \\
\nu=5 \end{array} \
\left[ \begin{array}{ccc}
0 & 0 & 0 \\
0 & 0 & 1 \\
0 & 1 & 0 \\
0 & 1 & 1 \\
1 & 0 & * \\ 
1 & 1 & 0 \\ 
1 & 1 & 1
\end{array} \right]
\begin{array}{l} 4 \\ 4 \\ 3,1 \\
\nu=4 \end{array} \
\left[ \begin{array}{ccc}
0 & 0 & 0 \\
0 & 0 & 1 \\
0 & 1 & 0 \\
0 & 1 & 1 \\
1 & 0 & z \\ 
1 & 0 & z^\perp \\ 
1 & 1 & z \\ 
1 & 1 & z^\perp
\end{array} \right]
\begin{array}{l} 4 \\ 4 \\ 2^2 \\
\nu=4 \end{array} \\ 
\left[ \begin{array}{ccc}
0 & 0 & 0 \\
0 & 0 & 1 \\
0 & 1 & z \\
0 & 1 & z^\perp \\
1 & 0 & z \\
1 & 0 & z^\perp \\
1 & 1 & 0 \\
1 & 1 & 1 
\end{array} \right]
\begin{array}{l} 4 \\ 4 \\ 2^2 \\
\nu=4 \end{array} \ 
\left[ \begin{array}{ccc}
0 & 0 & 0 \\
0 & 0 & 1 \\
0 & 1 & 0 \\
0 & 1 & 1 \\
1 & 0 & 0 \\ 
1 & 0 & 1 \\ 
1 & 1 & 0 \\ 
1 & 1 & 1
\end{array} \right]
\begin{array}{l} 4 \\ 4 \\ 4 \\
\nu=3 \end{array}
\end{eqnarray*}

\end{corollary}

\section{Classification of four-qubit orthogonal product bases}
\label{klasifikacija}

In this section we classify the four-qubit OPBs by using the 
weak equivalence defined in the previous section. 
More precisely, we solve our combinatorial problem in 
the case of four-qubits proposed in Sec. \ref{sec:combinatorial}. 
We obtain 33 equivalence classes of matrices in $\cO(4)$ and list 
the representatives $A$ of these classes. The corresponding 33 
families $\cF_A$ cover all OPBs up to equivalence. 
Equivalently, each OPB is equivalent to one belonging to these 
33 families. However, for a given OPB, such family $\cF_A$ does
not have to be unique. To obtain uniqueness, one has to replace 
the families $\cF_A$ by somewhat smaller families which we 
denote by $\cF_A^\#$ and call strict families. For this 
see section \ref{diskusija}.

\subsection{Preliminaries}
\label{subsec:pre}

We introduce two important lemmas. They will be used in proving the main result of this paper, Theorem \ref{thm:4qubit}. 

\bl \label{le:3q-2rows}
Let $A=[a_{i,j}]\in\cO(3)$ and assume that $a_{1,j}=a_{2,j}$ 
holds true for at most one $j\in\{1,2,3\}$.
If a row of vector variables 
$r=[~u~v~w~]$ is orthogonal to the last 6 rows of $A$, then 
$r$ is equal to the first or second row of $A$.
\el

\bpf
Let us assign to all vector variables (including $u$, $v$ and 
$w$) unit vectors such that different vector varables are 
assigned different unit vectors. It is understood that to a pair 
$x,x^\perp$ we assign a pair of orthogonal unit vectors. 
Let $\ket{\alpha_i}$ be the product vector obtained from the 
row $i$ of $A$, and $\ket{\xi}$ the product vector obtained 
from $r$. The hypothesis implies that the 2-dimensional 
subspace spanned by $\ket{\alpha_1}$ and $\ket{\alpha_2}$ 
contains no other product vectors (up to scalar multiple). 
As $\ket{\xi}$ belongs to this subspace, we must have 
$\ket{\xi}=\ket{\alpha_1}$ or $\ket{\xi}=\ket{\alpha_2}$. 
We conclude that $r$ must be equal to the first or second row 
of $A$.
\epf

By inspection of the matrix \eqref{eq:3qubit=8row}, it is easy to see that the following corollary holds.

\bcr
\label{cr:ir=1,3}
Let $A\in\cO(3)$ be irreducible and let $[~a~b~c]$ and $[~x~y~z]$ 
be two of its rows. Then the following assertions hold:

(i) each entry has multiplicity 1 or 3;

(ii) if two of the equalities $a=x^\perp$, $b=y^\perp$, 
$c=z^\perp$ hold, then all of them hold;

(iii) if $a=x^\perp$ and $\mu(a)=1$ then $b=y$ and $c=z$.
\ecr

These simple facts will be used many 
times in the proofs below and in the next section. 

For convenience we introduce some additional notation.
For any matrix $A\in\cO(4)$ and index sequences $i_1<\cdots<i_s$ and $j_1<\cdots<j_t$ we denote by 
$A[i_1,\ldots,i_s;j_1,\ldots,j_t]$ the $s\times t$ submatrix 
of $A$ contained in rows $i_1,\ldots,i_s$ and columns 
$j_1,\ldots,j_t$. 

Let $\nu_1=\nu_1(A)$, the maximal number of independent vector variables in the first column of $A$. We select $\nu_1$ 
independent varables $\xi_1,\xi_2,\ldots,\xi_{\nu_1}$ from 
this column and arrange them so that their multiplicities 
$\mu_{1,i}$ weakly decrease, i.e., $\mu_{1,1}\ge\mu_{1,2}\ge\cdots\ge\mu_{1,\nu_1}$. 
After permuting the rows of $A$, we may assume that its 
first column consists of $\mu_{1,1}$ entries $\xi_1$, 
followed by $\mu_{1,2}$ entries $\xi_2$,...,  then 
$\mu_{1,1}$ entries $\xi_1^\perp$,  
followed by $\mu_{1,2}$ entries $\xi_2^\perp$,.... 
Let us partition $A$ horizontally into $2\nu_1$ blocks 
$\tilde{N}_i$, $i=0,1,\ldots,2\n_1-1$ such that the elements in 
the first column of $\tilde{N}_i$ are all equal to 
$\xi_{i+1}$ if $i<\nu_1$ 
and are equal to $\xi_{i-\nu_1+1}^\perp$ if $i\ge\nu_1$. 
We let $N_i$ be the matrix obtained from 
$\tilde{N}_i$ by deleting the first column. 

Finally we define $M_i$ to be the matrix 
\begin{equation} \label{eq:Mat-M}
M_i:=\left[ \begin{array}{c}
N_i \\ N_{i+1} \\ \vdots \\ N_{i+\nu_1-1}
\end{array} \right],
\end{equation}
where the indices are to be reduced modulo $2\nu_1$.
Note that the rows of $M_i$ are mutually orthogonal, and so 
each of the matrices $M_i$ belongs to $\cO(3)$. 

In the case $\nu_1=1$ we just have $M_0=N_0$ and $M_1=N_1$.
Moreover, if $A$ is maximal then $M_0$ and $M_1$ are maximal 
and they have no vector variable in common.

In the next lemma we investigate the maximum of multiplicities of entries in the matrices of  $\cO(4)$. The symbols  
$M_i,N_i$ have been introduced above and we recall 
that $\mu_i=\mu_{i,1}$ is the largest of the $\mu_{i,j}$.

\bl
\label{le:mu>5}
If $A:=[ a_{s,i} ]\in\cO(4)$ and $\mu_1\ge\mu_2,\mu_3,\mu_4$ 
then $\mu_1\ge6$.
\el
\bpf
If $B<A$ it is immediate from the definition of "<" that 
$\mu_1(A)\le\mu_1(B)$. It follows that $\mu_1$ attains its minimum at a maximal matrix. Hence, without any loss of generality we can assume that $A$ is maximal. 

By Proposition \ref{pp:OPB} (iii) we have 
$\sum_{i=1}^4 \mu(a_{s,i})\ge 15$ for each 
$s\in\{1,2,\ldots,16\}$. Hence $\mu_1\ge4$. 

Suppose that $\mu_1=4$. Then $\mu(x)\in\{3,4\}$ for each entry 
$x$ of $A$. Since the multiplicities of any maximal set of independent vector variables of any column must add up to 8, we conclude that $\mu(x)=4$ for each entry $x$ of $A$. Thus each column of $A$ contains exactly two independent vector variables.
By permuting the rows of $A$, we may assume that the first column of $A$ is 
$[~a~a~a~a~b~b~b~b~a^\perp~a^\perp~a^\perp~a^\perp
~b^\perp~b^\perp~b^\perp~b^\perp~]^T$.
In this case all matrices $N_i$, $i=0,1,2,3$, have size 
$4\times3$. 

Suppose that at least one of the $M_i$, say $M_0$, is reducible. 
Thus one of the columns of $M_0$ contains only one independent 
variable, say $c$. By permuting the columns 2,3,4 of $A$ we may 
assume that the first column of $M_0$ has four of its entries equal to $c$ and the remaining four equal to $c^\perp$. As $\m(x)=4$ for all $x$, the first column of $M_2$ has four entries equal to $d$ and four equal to $d^\perp$ ($d$ another vector variable). Hence, $M_2$ is reducible. 
Since $M_1\in\cO(3)$, exactly 2 entries of $N_1$ are 
equal to $c$, and exactly 2 entries of $N_2$ are equal to $d$.
Consequently, we may assume that the second column of $A$ is
$[~c~c~c^\perp~c^\perp~c~c~c^\perp~c^\perp~d~d~d^\perp~
d^\perp~d~d~d^\perp~d^\perp~]^T$. 
Let us partition the $16\times2$ submatrix of $A$, consisting 
of the last two columns, into eight $2\times2$ blocks 
$L_i$, $i=0,1,\ldots,7$.  Note that the eight submatrices 
$K_i:=\left[ \begin{array}{c} L_i\\ L_{i+2} \end{array} \right]$, 
$i=0,1,\ldots,7$ (indices are modulo $8$) belong to $\cO(2)$. 
Moreover, the matrix
$K':=\left[ \begin{array}{c} L_3\\ L_4 \end{array} \right]$ also belongs to $\cO(2)$. 

Assume that one of the columns of some $L_i$ has two equal 
entries, say $x:=a_{1,3}=a_{2,3}$. 
Since $K_0\in\cO(2)$ we must have $a_{5,3}=a_{6,3}=x^\perp$. 
Similarly, since $K_2,K_4\in\cO(2)$, we deduce that $a_{9,3}=a_{10,3}=x$ and $a_{13,3}=a_{14,3}=x^\perp$. Since 
$K'\in\cO(2)$, we obtain that $a_{7,3}=a_{8,3}=x^\perp$. This is impossible since $\mu(x^\perp)=4$. We conclude that no column of any $L_i$ consists of two equal entries. 

Since the rows of $L_0$ are orthogonal, one of its columns, say the first, has the form $[~x~x^\perp~]^T$. 
Recall that each matrix in $\cO(2)$ is equivalent to one of 
the two matrices in \eqref{mat:2-q}. By inspection of these two matrices and by taking into account that the two entries of the second column of $L_0$ are not equal, we deduce that the first 
column of $L_2$ must consist of $x$ and $x^\perp$. 
By repeating this argument, it follows that the entries of 
the first column of $L_i$ for $i$ even are $x$ and $x^\perp$. 
As $K'\in\cO(2)$, the same is true for $L_3$. 
This is impossible since $\mu(x)=4$. 
We conclude that all $M_i$ are irreducible.

As $M_0$ is irreducible, there are independent vector variables 
$x,y,z,u,v,w$ such that $x,y,z$ have multiplicity 3 in $M_0$ and
$u,v,w$ multiplicity 1 in $M_0$, with $x,u$ in the first column,
$y,v$ in the second and $z,w$ in the third column of $M_0$. 
Consequently, $x,y,z$ have multiplicity 1 in $M_2$ and $u,v,w$ 
multiplicity 3 in $M_2$. As $M_2$ is also irreducible, by using 
Example \ref{ex:2i3kubita}, we infer that there exist 
permutation matrices $P$ and $Q$ such that
\bea
M_0=P \left[ \begin{array}{ccc}
x & y & z \\
u & y^\perp & z \\
u^\perp & y^\perp & z \\
x & v & z^\perp \\
x & v^\perp & z^\perp \\
x^\perp & y & w \\
x^\perp & y & w^\perp \\
x^\perp & y^\perp & z^\perp
\end{array} \right], \qquad
M_2=Q \left[ \begin{array}{ccc}
u & v & w \\
x & v^\perp & w \\
x^\perp & v^\perp & w \\
u & y & w^\perp \\
u & y^\perp & w^\perp \\
u^\perp & v & z \\
u^\perp & v & z^\perp \\
u^\perp & v^\perp & w^\perp 
\end{array} \right].
\eea

Let $R:=[s^\perp~u~v~w]$, $s\in\{a,b\}$, be the row 
of $A$ containing the row $[u~v~w]$ of $M_2$. 
Let $t\in\{a,b\}$ be different from $s$. 
Since $[u~v~w]$ is orthogonal to only three rows of $M_0$, 
$R$ is not orthogonal to at least one of the four rows of $A$ whose first element is $t$. Thus we have a contradiction.
We conclude that $\mu_1\ge5$.

Suppose that $\mu_1=5$. There are three possibilities for the 
partition $\pi_1$ associated to the first column of $A$. 
In each of these cases we shall obtain a contradiction.

Case 1: $\pi_1=5,3$. 

We may assume that
$[~a~a~a~a~a~b~b~b~a^\perp~a^\perp~a^\perp~a^\perp
~a^\perp~b^\perp~b^\perp~b^\perp~]^T$ 
is the first column of $A$. 
In this case $N_0$ and $N_2$ have the size $5\times3$, while 
$N_1$ and $N_3$ have size $3\times3$.
Assume that one of the columns of $N_1$, say the first column, 
consists of 3 equal entries $x=a_{6,2}=a_{7,2}=a_{8,2}$. 
Since $M_0\in\cO(3)$, the first column of $N_0$ must contain 
at least 3 entries $x^\perp$. Similarly, the first column of 
$N_2$ must contain at least 3 entries $x^\perp$. This 
contradicts the inequality $\mu(x^\perp)\le5$. 
We conclude that no column of $N_1$ consists of 3 equal entries.

Subcase 1a: At least one of the $M_i$, say $M_0$, is reducible. 

We may assume that $a_{i,2}\in\{c,c^\perp\}$ for $1\le i\le8$.
As the three entries of the first column of $N_1$ are not equal, we may assume that the first column of $M_0$ is 
$[~c^\perp~c^\perp~c^\perp~c~c~c~c~c^\perp~]^T$.  
Since $M_3\in\cO(3)$, one of the entries in 
the first column of $N_3$ must be $c$. 
As $\mu_1=5$ we must have $\mu(c)=5$. 
Hence, we may assume that the first column of $N_3$ is 
$[~c~d~d^\perp~]$, where $d\not\in\{c,c^\perp\}$. 
Note that $c$ has multiplicity $2$ in $M_1$. 
By Corollary \ref{cr:ir=1,3} (i), $M_1$ is reducible. Thus, we may assume that $a_{i,3}\in\{e,e^\perp\}$ for $6\le i\le13$, where $e$ is a new vector variable. 
Since the entries in the second column of $N_1$ cannot be the same, we may assume that the second column of $N_1$ has two entries $e$ and one entry $e^\perp$. 
Since the rows 4,5,6,7 of $A$ are orthogonal to each other, 
we must have $A[4,5,6,7;3,4]\in\cO(2)$.  
The first two entries of the second column of $N_1$ cannot be both equal to $e$. Indeed, this would imply that $a_{4,3}=a_{5,3}=e^\perp$ which contradicts the 
fact that $e^\perp$ has multiplicity 4 in $M_1\in\cO(3)$.
Consequently, we may assume that the second column of $N_1$ is $[~e~e^\perp~e~]^T$. 
Since $M_0\in\cO(3)$ and $\mu(e)\le5$, we infer that $e^\perp$ 
occurs exactly once in $N_0$. Similarly, $e$ occurs exactly once in $N_3$. We may assume that it does not occur in the 
last row of $A$. 
The first 3 entries of this row have multiplicity at most 3. 
Hence the last entry must have multiplicity at least 6 by Proposition \ref{pp:OPB} (iii). We have a contradiction with the assumption $\m_1=5$.

Subcase 1b: Each $M_i$ is irreducible.

Thus $M_0$ is equivalent to the matrix \eqref{eq:3qubit=8row}.
Suppose that the entries of $N_1$ are pairwise distinct.
It is easy to check that any $3\times3$ submatrix of the 
matrix \eqref{eq:3qubit=8row}, which has pairwise distinct 
entries, cannot include the first or last row and must 
contain only one row of each pair of rows $\{2,3\}$, 
$\{4,5\}$, $\{6,7\}$. Hence, we may assume that 
$$
N_1 = \left[ \begin{array}{ccc}
u & y & z \\
x & v & z^\perp \\
x^\perp & y^\perp & w
\end{array} \right].
$$
The entries $u,v,w$ in $N_1$ correspond to $a$ or $a^\perp$, 
$b$ or $b^\perp$, $c$ or $c^\perp$ in the matrix \eqref{eq:3qubit=8row}, respectively. Hence, each of $x,y,z$ must have multiplicity 3 in $M_0$. Similarly, they also have 
multiplicity 3 in $M_1$. Since $M_3$ is irreducible, the first column of $N_3$ must contain the elements $u,x,x^\perp$. 
Thus $x$ occurs 2 times in both $N_0$ and $N_2$ and once in 
both $N_1$ and $N_3$. This contradicts the inequality $\mu(x)\le5$. We conclude that at least one of the columns of $N_1$ must contain two equal entries. 

We may assume that $c:=a_{6,2}=a_{7,2}$. Since $M_0$ and $M_1$ 
are irreducible, the multiplicity of $c$ in $M_0$ and $M_1$ 
must be 3. If $a_{8,2}\ne c^\perp$ then both $N_0$ and $N_2$ 
would contain 3 entries $c^\perp$, contradicting the 
inequality $\mu(c)\le5$. Thus we must have  $a_{8,2}=c^\perp$.
It follows that both $N_0$ and $N_2$ contain exactly two entries 
equal to $c^\perp$ and only one entry equal to $c$. 
Since $M_2$ is irreducible, the multiplicity  of $c^\perp$ in 
$M_2$ must be 3. It follows that the first column of $N_3$ 
has two entries equal to $c$. This contradicts the inequality $\mu(c)\le5$.

Case 2: $\pi_1=5,2,1$.  

We may assume that
$[~a~a~a~a~a~b~b~c~a^\perp~a^\perp~a^\perp~a^\perp
~a^\perp~b^\perp~b^\perp~c^\perp~]^T$ 
is the first column of $A$. Since both $M_0,M_5\in\cO(3)$, 
Lemma \ref{le:1-row} implies that $N_2=N_5$. 
Let $N_2=[~x~y~z~]$. 
As $\mu(c)=1$, Proposition \ref{pp:OPB} (iii) implies that 
$\mu(x)+\mu(y)+\mu(z)\ge14$. We may assume that 
$\mu(x)=\mu(y)5$ and $\mu(z)=4$. 
Thus $x$ must have multiplicity 2 or 4 in either $M_0$ or $M_3$. 
By Corollary \ref{cr:ir=1,3} (i), either $M_0$ or $M_3$ is reducible. We may assume that $M_0$ is reducible. 
Since $\mu(z)=4$ and $a_{8,4}=a_{16,4}=z$, either $x$ or $y$ must have multiplicity 4 in $M_0$. We may assume that $x$ 
has multiplicity 4 in $M_0$. Then $x$ occurs only once in $M_3$, and so $x$ occurs in neither $N_3$ nor $N_4$. 
Since $M_1\in\cO(3)$ and $a_{6,2},a_{7,2}\in\{x,x^\perp\}$, 
the multiplicity of $x$ in $M_1$ must be 2. Moreover, 
$a_{6,2}\ne a_{7,2}$. Indeed, $a_{6,2}=a_{7,2}=x$ contradicts the fact that $x$ has multiplicity 2 in $M_1$ and  
$a_{6,2}=a_{7,2}=x^\perp$ contradicts the fact that $x$ does 
not occur in $N_3$. 
Furthermore we cannot have $a_{6,3}=a_{7,3}$ and 
$a_{6,4}=a_{7,4}$ since $A$ is maximal and the submatrix 
$A[6,7;2]$ would contradict Proposition \ref{pp:sused} (iii). 
Hence we can now apply Lemma \ref{le:3q-2rows} to $M_1$ and 
the two rows of $N_4$. We deduce that the two entries 
in the first column of $N_4$ must belong to $\{x,x^\perp\}$.
This contradicts the fact that $\mu(x)=5$.

Case 3: $\pi_1=5,1^3$. We may assume that
$[~a~a~a~a~a~b~c~d~a^\perp~a^\perp~a^\perp~a^\perp
~a^\perp~b^\perp~c^\perp~d^\perp~]^T$ 
is the first column of $A$. 
Since both $M_0,M_7\in\cO(3)$, Lemma \ref{le:1-row} implies that $N_3=N_7$. Let $N_3=[x~y~z]$. Similarly, we can show that $N_2=N_6$ and $N_1=N_5$. 
As in the previous case, we can show that $M_0$ or $M_4$ is reducible. We may assume that $M_0$ is reducible, and that $x$ has multiplicity 4 in $M_0$. 
Since $N_i=N_{i+4}$ for $i=1,2,3$, we must have $\mu(x)>5$ or $\mu(x^\perp)>5$. Thus we have a contradiction.

This completes the proof.
\epf

\subsection{Main result} 
\label{sec:proof}

Let us now state our main result. 
\bt
\label{thm:4qubit}
The maximal matrices in $\cO(4)$ split into 33 equivalence 
classes. They are grouped into 15 switching classes whose 
representatives are listed below (some columns are normalized). 
When a switching class contains more than one equivalence class, 
we list their representatives in the appendix.
\et

\bea \label{eq:sw1-3} 
&& \left[ \begin{array}{cccc}
0 & 0 & 0 & * \\
0 & 0 & 1 & * \\
0 & 1 & c & * \\
0 & 1 & c^\perp & * \\
1 & b & x & * \\
1 & b & x^\perp & * \\
1 & b^\perp & y & * \\
1 & b^\perp & y^\perp & * \\  
\end{array} \right]
\begin{array}{l}8 \\ 4^2 \\ 2^4 \\ 1^8 \\
\nu=15 \end{array}\quad
\left[ \begin{array}{cccc}0 & 0 & c & * \\
0 & 0 & c^\perp & * \\
0 & 1 & x & * \\
0 & 1 & x^\perp & * \\
1 & b & 0 & 0 \\
1 & * & 1 & 0 \\
1 & b & * & 1 \\
1 & b^\perp & 0 & * \\
1 & b^\perp & 1 & 1 \\
\end{array} \right]
\begin{array}{l} 8 \\ 4,3,1 \\ 3,2^2,1 \\ 3,1^5 \\
\nu=14 \end{array}\quad
\left[ \begin{array}{cccc}
0 & 0 & 0 & * \\
0 & b & 1 & * \\
0 & b^\perp & 1 & * \\
1 & 0 & c & * \\
1 & 0 & c^\perp & * \\
1 & 1 & 1 & * \\
a & 1 & 0 & * \\
a^\perp & 1 & 0 & * \\
\end{array} \right]
\begin{array}{l} 6,2 \\ 6,2 \\ 6,2 \\ 1^8 \\
\nu=14 \end{array}\quad
\eea

\bea \label{eq:sw4-6}
&& \left[ \begin{array}{cccc}
0 & 0 & 0 & 0 \\
0 & * & 1 & 0 \\
0 & 0 & * & 1 \\
0 & 1 & 0 & * \\
0 & 1 & 1 & 1 \\
1 & b & c & d \\
1 & * & c^\perp & d \\
1 & b & * & d^\perp \\
1 & b^\perp & c & * \\
1 & b^\perp & c^\perp & d^\perp \\
\end{array} \right]
\begin{array}{l} 8 \\3^2,1^2\\ 3^2,1^2\\ 3^2,1^2 \\
\nu=13 \end{array}
\left[ \begin{array}{cccc}
0 & 0 & 0 & * \\
0 & 1 & 0 & 1 \\
0 & b & 1 & * \\
0 & b^\perp & 1 & * \\
1 & 1 & 1 & 0 \\
1 & 1 & * & 1 \\
1 & 0 & c & * \\
1 & 0 & c^\perp & * \\
* & 1 & 0 & 0 \\
\end{array} \right]
\begin{array}{l} 7,1 \\ 6,2 \\ 5,2,1 \\ 3,1^5 \\
\nu=13 \end{array}\quad
\left[ \begin{array}{cccc}
0 & 0 & 0 & 0 \\
0 & * & 0 & 1 \\
0 & b & 1 & * \\
0 & b^\perp & 1 & * \\
1 & 1 & 1 & 0 \\
1 & 1 & * & 1 \\
1 & 0 & c & * \\
1 & 0 & c^\perp & * \\
* & 1 & 0 & 0 \\
\end{array} \right] 
\begin{array}{l} 7,1 \\ 5,2,1 \\ 5,2,1 \\ 4,1^4 \\
\nu=13 \end{array}\quad
\eea

\bea \label{eq:sw7-9}
&& \left[ \begin{array}{cccc}
0 & 0 & 0 & 0 \\
0 & 0 & 1 & * \\
0 & 1 & c & * \\
0 & 1 & c^\perp & * \\
1 & 1 & 1 & 0 \\
1 & 1 & 0 & * \\
1 & 0 & * & 0 \\
1 & * & 1 & 1 \\
* & 0 & 0 & 1 \\
\end{array} \right]
\begin{array}{l} 7,1 \\ 7,1 \\ 5,2,1 \\ 4,1^4 \\
\nu=12 \end{array}
\left[\begin{array}{cccc}
0 & 0 & 0 & 0 \\
0 & 1 & 0 & d \\
0 & * & 1 & d \\
0 & 1 & * & d^\perp \\
0 & 0 & 1 & d^\perp \\
1 & 1 & c & * \\
1 & 1 & c^\perp & * \\
1 & 0 & 1 & 1 \\
1 & 0 & * & 0 \\
* & 0 & 0 & 1 \\
\end{array} \right]
\begin{array}{l} 7,1 \\ 7,1 \\ 4,2,1^2 \\ 3^2,1^2 \\
\nu=12 \end{array}\quad
\left[ \begin{array}{cccc}
0 & b & 0 & 1 \\
0 & 0 & 0 & 0 \\
0 & b & 1 & * \\
0 &b^\perp & * & 1 \\
0 &b^\perp & 1 & 0 \\
1 & 1 & 0 & 1 \\
1 & 1 & 1 & * \\
1 & 0 & * & d \\
1 & 0 & * & d^\perp \\
* & 1 & 0 & 0 \\
\end{array} \right]
\begin{array}{l} 7,1 \\ 5,3 \\ 5,1^3 \\ 4,2,1^2 \\
\nu=12 \end{array}\quad
\eea

\bea \label{eq:sw10-12}
&& \left[ \begin{array}{cccc}
0 & 0 & 0 & 0 \\
0 & 1 & 1 & 1 \\
0 & * & 1 & 0 \\
0 & 0 & * & 1 \\
1 & 1 & 1 & d \\
1 & 0 & 0 & d^\perp \\
1 & * & 1 & d^\perp \\
1 & 0 & * & d \\
a & 1 & 0 & * \\
a^\perp & 1 & 0 & * \\
\end{array} \right]
\begin{array}{l} 6,2 \\ 6,1^2 \\ 6,1^2 \\ 3^2,1^2 \\
\nu=12 \end{array} \quad
\left[ \begin{array}{cccc}
0 & * & 1 & 0 \\
0 & 0 & * & 1 \\
0 & 1 & 0 & * \\
1 & * & 0 & 1 \\
1 & 1 & * & 0 \\
1 & 0 & 1 & * \\
* & 0 & 0 & 0 \\
* & 1 & 1 & 1 \\
\end{array} \right] 
\begin{array}{l} 6,1^2 \\ 6,1^2 \\ 6,1^2 \\ 6,1^2 \\
\nu=12 \end{array}\quad
\left[ \begin{array}{cccc}
0 & 0 & 0 & 0 \\
0 & * & 1 & 0 \\
0 & 0 & * & 1 \\
0 & 1 & 0 & * \\
1 & 1 & 1 & 0 \\
1 & 0 & 1 & d \\
1 & * & 0 & d \\
1 & 0 & * & d^\perp \\
1 & 1 & 0 & d^\perp \\
* & 1 & 1 & 1 \\
\end{array} \right]
\begin{array}{l} 7,1 \\ 6,1^2 \\ 6,1^2 \\ 4,3,1 
\\ \nu=11 \end{array}\quad
\eea


\bea \label{eq:sw13-15}
&& \left[ \begin{array}{cccc}
0 & 0 & 0 & 0 \\
0 & 1 & 0 & d \\
0 & * & 1 & d \\
0 & 1 & * & d^\perp \\
0 & 0 & 1 & d^\perp \\
1 & 0 & 1  & 1\\
1 & 0 & c & 0\\
1 & 1 & c & *\\
1 & * & c^\perp & 0 \\
1 & 1 & c^\perp & 1 \\
* & 0 & 0 & 1 \\
\end{array} \right]
\begin{array}{l} 7,1 \\ 6,1^2 \\ 4,3,1 \\ 4,3,1 \\
\nu=11 \end{array}\quad
\left[ \begin{array}{cccc}
0 & 0 & 0 & 0 \\
0 & 1 & 1 & d \\
0 & 0 & * & 1 \\
0 & 1 & * & d^\perp \\
1 & 1 & 1 & 0 \\
1 & 0 & 0 & d \\
1 & * & 0 & d^\perp \\
1 & * & 1 & 1 \\
* & 0 & 1 & 0 \\
* & 1 & 0 & d \\
\end{array} \right]
\begin{array}{l} 6,1^2 \\ 6,1^2 \\ 6,1^2 \\ 4^2 \\
\nu=11 \end{array}\quad
\left[ \begin{array}{cccc}
0 & 0 & 0 & 0 \\
0 & 0 & c & 1 \\
0 & 1 & 0 & d \\
0 & * & 1 & 0 \\
0 & 1 & 1 & 1 \\
1 & 0 & c^\perp & 0 \\
1 & 0 & c & d^\perp \\
1 & 1 & c^\perp & d \\
1 & * & c & d \\
1 & 1 & 1 & d^\perp \\
* & 0 & c^\perp & 1 \\
* & 1 & 0 & d^\perp \\
\end{array} \right]
\begin{array}{l} 6,1^2 \\ 6,1^2 \\ 4^2 \\ 4^2 \\
\nu=10 \end{array}
\eea

These matrices are arranged so that the parameter $\nu$ 
decreases from $\nu=15$ to $\nu=10$. For fixed value of $\nu$, 
the matrices are listed in decreasing lexicographic 
order of the partitions $\pi_1,\pi_2,\pi_3,\pi_4$.

Strictly speaking, the above matrices are not members 
of $\cO(4)$ because some columns of these matrices 
are normalized. We normalize column $j$ if $\mu_j>1$ and
we choose the normalization so that $\mu(0)=\mu_j$. 
To get the genuine representatives one has 
to replace in each column the entries 0 and 1 with 
a new vector variable and its perpendicular, respectively. 
Of course different variables have to be 
used for different columns. 
When counting the number of independent variables one has 
to undo the normalization. 
We have arranged the columns so that
$\pi_1\ge \pi_2\ge \pi_3\ge \pi_4$. After each representative $A$ we show the associated partitions $\pi_i$ and the parameter 
$\nu=\nu_A$. 

For instance let us consider the first matrix in \eqref{eq:sw1-3}. After undoing the normalization, we obtain 
the matrix
$$
A:= \left[ \begin{array}{cccc}
u & v & w & * \\
u & v & w^\perp & * \\
u & v^\perp & c & * \\
u & v^\perp & c^\perp & * \\
u^\perp & b & x & * \\
u^\perp & b & x^\perp & * \\
u^\perp & b^\perp & y & * \\
u^\perp & b^\perp & y^\perp & * \\  
\end{array} \right],
$$
where we used new independent variables $u,v,w$. 
In this matrix each of the 8 rows really stands for two rows 
because each entry in the last column is an asterisk.
The first column of this matrix contains only one independent 
variable, say $u$, and its multiplicity is 8. Consequently, 
the first partition is $\pi_1=8$. The second column has two 
independent variables, say $v$ and $b$. Each of them has multiplicity 4, and so $\pi_2=4^2$. Similarly for the third
and fourth columns we obtain the partitions $\pi_3=2^4$ and 
$\pi_4=1^8$. The largest parts of these partitions are
$\mu_1=8$, $\mu_2=4$, $\mu_3=2$, and $\mu_4=1$, respectively. 
Since $\nu_i$ is the number of parts of the partition 
$\pi_i$, we have $\nu_1=1$, $\nu_2=2$, $\nu_3=4$, $\nu_4=8$
and so $\nu_A=15$. 

We number the switching classes in the order that they are 
listed in \eqref{eq:sw1-3}-\eqref{eq:sw13-15}. 
The  classes 1,2,4 consist of reducible and the other of irreducible matrices. 
Each of the switching classes 4,11,12,13,14,15 consists of  just one equivalence class. Each of the other nine switching classes contains at least two  equivalence classes. The representatives 
of these equivalence classes are listed in the appendix. In total there are 33 equivalence classes.  

Next we prove our main result, Theorem  \ref{thm:4qubit}. 

\bpf
We shall first prove that our list of representatives of the 
switching classes of maximal matrices of $\cO(4)$ is complete. 
In other words, we have to show that any maximal matrix 
$A=[a_{i,j}]\in\cO(4)$ is weakly equivalent to one of 
the matrices listed in \eqref{eq:sw1-3}-\eqref{eq:sw13-15}. 

In this proof, we shall use the notation $\mu_i,M_i,N_i$ introduced just before Lemma \ref{le:mu>5}. Recall that 
each of the submatrices $M_i$ of $A$ belongs to $\cO(3)$.

Denote by $\pi_j$ the partition associated to the column $j$ 
of $A$. By permuting the columns, we may assume that 
$\pi_1\ge\pi_2,\pi_3,\pi_4$. In spite of this condition, the partition $\pi_1$ may vary over a given switching class. 
Our representatives (as listed in the theorem) are chosen so that $\pi_1$ is maximal over all matrices in its switching class.

In view of Lemma \ref{le:mu>5}, we have $\mu_1\in\{6,7,8\}$.
We divide the proof into four cases according to the partition 
$\pi_1$ associated to the first column of $A$. 
In each of these four cases we assume that the matrix $A$ is a representative of some switching class (in particular, $A$ is 
maximal) and that it is chosen so that the partition $\pi_1$ is maximal. If during the proof it turns out that $A$ is weakly equivalent to a matrix having bigger partition $\pi_1$, then we can discard such $A$.

Case 1: $\pi_1=8$. 

Since $A$ is maximal and reducible, both matrices $M_0=N_0$ and $M_1=N_1$ must be maximal in $\cO(3)$, and moreover no vector variable occurs in $M_0$ and $M_1$. There are three subcases for the submatrices $M_0$ and $M_1$ as follows. 

\noindent 1a: they are both reducible; \\
\noindent 1b: one of them is reducible and the other irreducible; \\
\noindent 1c: they are both irreducible. 

Note that the first 8 rows of $A$ as well as the last 8 rows of $A$ are $\{2,3,4\}$-compatible. Hence we can permute arbitrarily 
the rows of $M_0$ as well as those of $M_1$. We can also 
permute arbitrarily and independently the columns of $M_0$ and 
the columns of $M_1$ because these operations are switching operations. 
By using Lemma \ref{le:3qubits}, we infer that each of these three subcases gives a single switching class in $\cO(4)$. 
These are the switching classes 1,2,4 respectively.

Case 2: $\pi_1=7,1$. 

By permuting the rows of $A$, we may assume that 
$[~a~a~a~a~a~a~a~b~a^\perp~a^\perp~a^\perp~a^\perp
~a^\perp~a^\perp~a^\perp~b^\perp~]^T$ 
is the first column of $A$. 
Since both $M_0,M_3\in\cO(3)$, Lemma \ref{le:1-row} implies that 
$N_1=N_3$. 
Set $x:=a_{8,2}=a_{16,2}$, $y:=a_{8,3}=a_{16,3}$ and 
$z:=a_{8,4}=a_{16,4}$. 

Subcase 2a: Some $M_i$, say $M_0$, is reducible. 

By permuting the first 7 rows of $A$, we may assume that the first column of $M_0$ is 
$[~x^\perp~x^\perp~x^\perp~x^\perp~x~x~x~x~]^T$. 
By applying Proposition \ref{pp:sused} (iii) to the submatrix  $A[1,2,3,4;3,4]\in\cO(2)$, we infer that neither $y$ nor $z$ 
occurs in it. 
By using the switching operation of interchanging the two 
columns of $A[1,2,3,4;3,4]$, we can always assume that 
$a_{1,3}$ has multiplicity 2 in this submatrix, even after 
interchanging the last two columns of $A$. 
By permuting the first four rows of $N_0$, we may assume that 
$a_{1,3}=a_{2,3}=r$, 
$a_{3,3}=a_{4,3}=r^\perp$, 
$a_{1,4}=s$, $a_{2,4}=s^\perp$, 
$a_{3,4}=t$, $a_{4,4}=t^\perp$, 
where $r,s,t$ are independent variables which do not 
occur in $A$ outside the submatrix $A[1,2,3,4;3,4]$. 

Suppose that $M_2$ is reducible. 

Then $y$ or $z$ has multiplicity 4 in $M_2$. By interchanging 
the last two columns of $A$ (if necessary) we may assume that 
$y$ has multiplicity 4 in $M_2$. By permuting the rows of 
$N_2$, we may assume that $a_{i,3}$ 
is equal to $y^\perp$ for $i=9,10,11,12$ and to $y$ for 
$i=13,14,15$. By using weak equivalence, we can assume 
that $a_{9,2}$ has multiplicity 2 in 
$A[9,10,11,12;2,4]\in\cO(2)$. 
By applying Proposition \ref{pp:sused} to this 
submatrix and by permuting the first four rows of $N_2$, 
we may assume that $a_{9,2}=a_{10,2}=u$, 
$a_{11,2}=a_{12,2}=u^\perp$, 
$a_{9,4}=v$, $a_{10,4}=v^\perp$, 
$a_{11,4}=w$, $a_{12,4}=w^\perp$, 
where $u,v,w$ are independent variables which do not 
occur in $A$ outside the submatrix $A[9,10,11,12;2,4]$. 
Denote by $p$ and $q$ the multiplicity of $x$ and $y$ 
in $A[13,14,15,16;2,4]\in\cO(2)$ and 
$A[5,6,7,8;3,4]\in\cO(2)$, respectively, 
and note that $p,q\in\{1,2\}$.
If $p=q=1$ then $\pi_2=\pi_3=5,2,1$; $\pi_4=4,1^4$ and 
$A$ belongs to the switching class 6. Similarly, if $p\ne q$ then $A$ belongs to the switching class 5. 
If $p=q=2$ then we may assume that 
$a_ {7,3}=y$, $a_{7,4}=z^\perp$, $a_{15,2}=x$ and 
$a_{15,4}=z^\perp$. 
By introducing a new vector variable, say $e$, and 
setting $a_{7,1}=e$ and $a_{15,1}=e^\perp$, we obtain a new matrix in $\cO(4)$. This contradicts the maximality of $A$. 
This rules out the possibility $p=q=2$. 

Suppose now that $M_2$ is irreducible. 

If the multiplicity of $x$ in $M_2$ is 1, 
then we may assume that the first column of $M_2$ is 
$[~u~u~u~u^\perp~u^\perp~u^\perp~x^\perp~x~]$, 
and so $\pi_2=5,3$.
Since $M_2$ is equivalent to the matrix \eqref{eq:3qubit=8row}, 
we must have $a_{15,3}=y$ and $a_{15,4}=z$. Hence, both 
$y$ and $z$ have multiplicity 3 in $M_2$. 
By interchanging the two columns of $A[1,2,3,4;3,4]$ if 
necessary (a switching operation), we may assume that 
$a_{1,3}$ has multiplicity 2 in that submatrix. 
If the multiplicity of $y$ in $A[5,6,7,8;3,4]\in\cO(2)$ is 1  
then $\pi_3=4,2,1^2$ and $\pi_4=5,1^3$, and if it is 2 then $\pi_3=5,2,1$, $\pi_4=4,1^4$. In both case $A$ belongs to the switching class 9.  

If the multiplicity of $x$ in $M_2$ is 3, then we may assume that $a_{i,2}$ is equal to $x^\perp$ for $i=11,12,13$ and to $x$ 
for $i=14,15$. Thus $\pi_2=7,1$. Moreover, $u:=a_{9,3}=a_{10,3}$ 
and $v:=a_{9,4}=a_{10,4}$, and so both $u$ and $v$ 
have multiplicity 3 in $M_2$. As the rows 8 and 9 are 
orthogonal, we have $u=y^\perp$ or $v=z^\perp$. 
Note that it is impossible that both equalities hold.
By interchanging the last two columns of $A$ (and using the weak 
equivalence) we may assume that $u=y^\perp$ and $v\ne z^\perp$.

Suppose that $v=z$. Since $A[9,14,15,16;3,4]\in\cO(2)$, 
we must have $a_{14,4}=a_{15,4}=z^\perp$, and we may assume 
that $a_{11,4}=z^\perp$. Consequently, 
$a_{13,4}=a_{12,4}^\perp$. Since $M_2$ is irreducible, it follows that $a_{12,3}=a_{13,3}=y$ and $a_{11,3}=y^\perp$. 
As $A[5,6,7,8;3,4]\in\cO(2)$, one of $y,z$ must occur twice in 
this submatrix. If $y$ occurs twice, then we may assume that 
$a_{5,3}=a_{6,3}=y^\perp$, $a_{7,3}=y$ and $a_{7,4}=z^\perp$. 
Moreover, $w:=a_{5,4}=a_{6,4}^\perp$ with $w$ and $z$ independent. By interchanging the two columns of the submatrix 
$A[1,2,3,4;3,4]\in\cO(2)$ if necessary (a switching operation), 
we may assume that $a_{1,3}$ has multiplicity 2 in this submatrix. Hence, $\pi_3=5,2,1$, $\pi_4=4,1^4$ and $A$ belongs to the switching class 7. Similarly, if $z$ occurs twice in 
$A[5,6,7,8;3,4]$ we can verify that $A$ again belongs to the switching class 7.

Finally suppose that $v$ and $z$ are independent, and so $z$ 
occurs only once in $M_2$.
As $P:=A[10,11,12,13;3,4]\in\cO(2)$ and $z$ does not occur 
in $P$, we may assume that 
$a_{11,4}=v$, $a_{i,4}=v^\perp$ for $i=12,13,14$ and 
$a_{15,4}=z^\perp$. 
It follows now that $a_{11,3}=a_{15,3}=y$ and 
$a_{14,3}=y^\perp$.
If $y$ has multiplicity 2 in $Q:=A[5,6,7,8;3,4]\in\cO(2)$, 
then we may assume that $a_{7,3}=y$ and $a_{7,4}=z^\perp$. 
But then we can set $a_{7,1}=w$ and $a_{15,1}=w^\perp$, 
where $w$ is a new vector variable, to obtain a new matrix 
in $\cO(4)$. This contradicts the maximality of $A$. 
We conclude that $y$ has multiplicity 1 in $Q$ and
$z$ multiplicity 2. Hence $\pi_3=4,2,1^2$, $\pi_4=3^2,1^2$ 
and $A$ belongs to the switching class 8.

Subcase 2b:  All $M_i$ are irreducible. 

Let us introduce two submatrices $P:=A[2,3,4,5;3,4]$ and 
$Q:=A[10,11,12,13;3,4]$.
Note that each row of the matrix \eqref{eq:3qubit=8row} 
contains at most one entry of multiplicity 1. 
As $M_0$ is irreducible, it is equivalent to the matrix 
\eqref{eq:3qubit=8row} and so at least two of the entries 
$x,y,z$ must have multiplicity 3 in $M_0$. 
We may assume that this holds true for $x$ and $y$. 
For the same reason, at least one of $x,y$, say $x$, has multiplicity 3 in $M_2$ and so $\pi_2=6,1^2$.
We may assume that $a_{i,2}$ is equal to $x^\perp$ 
for $i=3,4,5,11,12,13$ and equal to $x$ for $i=6,7,14,15$. 
Consequently, $a_{1,3}=a_{2,3}$, $a_{1,4}=a_{2,4}$ and 
$v:=a_{9,3}=a_{10,3}$, $w:=a_{9,4}=a_{10,4}$. 
We infer that $a_{1,3}$ and $a_{1,4}$ have multiplicity 3 
in $M_0$, and $v$ and $w$ have multiplicity 3 in $M_2$. Since both $a_{1,3}$ and $y$ have multiplicity 3 in 
$M_0$, we must have $a_{1,3}\in\{y,y^\perp\}$. 
Note that the submatrices $P$ and $Q$ belong to $\cO(2)$.

Suppose that $a_{1,3}=y$. 

Since rows 1 and 8 are orthogonal, 
we have $a_{1,4}=a_{2,4}=z^\perp$. 
By permuting the rows 3,4,5 (if necessary), we may assume 
that $a_{3,3}=y^\perp$. It follows that $a_{3,4}=z^\perp$. 
Thus $z^\perp$ has multiplicity 2 in the submatrix $P$.
Therefore we have $a_{4,4}=a_{5,4}=z$. 
Now observe that we can apply Proposition \ref{pp:sused} (iii) to the submatrix  $A[4,5;3]\in\cO(1)$. 
We infer that the variable $a_{4,3}=a_{5,3}^\perp$ has multiplicity 1 in $A$.  
As we must have $a_{7,3}=a_{8,3}=y^\perp$, the same argument shows that the variable $a_{6,4}=a_{7,4}^\perp$ has multiplicity 1 in $A$.  
Since row 8 and 9 are orthogonal, $v=y^\perp$ or $w=z^\perp$. 
Note that we cannot have $v=y^\perp$ and $w=z^\perp$. 

We claim that $w\ne z$ and $v\ne y$.
If $w=z$ then $z$ has multiplicity 3 in $M_2$ and we may assume that $a_{11,4}=z^\perp$. 
By Corollary \ref{cr:ir=1,3}, applied to the rows 3 and 8 of 
$M_2$, we have $a_{11,3}=y^\perp$. 
By applying Proposition \ref{pp:sused} (iii) to the submatrix  $A[3,11;1]\in\cO(1)$, we obtain a contradiction. 
We conclude that $w\ne z$. Similarly, $v\ne y$. 
Thus our claim is proved.

Assume that $v=y^\perp$. If $y$ occurs twice in the submatrix 
$Q$, then we may assume that $a_{11,3}=y^\perp$.
By inspecting the matrix \eqref{eq:3qubit=8row}, we infer that 
$a_{11,4}=z^\perp$.
By applying Proposition \ref{pp:sused} (iii) to the submatrix  $A[3,11;1]\in\cO(1)$, we obtain a contradiction. 
We conclude that $y$ occurs only once in $Q$ and we may assume that $a_{11,3}=y$ and $a_{13,3}^\perp=a_{14,3}=y$. 
It follows easily that $a_{15,4}=z^\perp$, $a_{14,4}=w^\perp$, 
$a_{12,4}=a_{13,4}=w^\perp$ and $a_{11,3}=w$. 
We have $\pi_3=6,1^2$ and $\pi_4=4,3,1$. This $A$ belongs to 
the switching class 12.

Assume now that $v$ and $y$ are independent. Then we must have 
$w=z^\perp$. Since $y$ occurs only once in $M_2$, we may 
assume that $a_{15,3}=y^\perp$. It follows that $v$ has 
multiplicity 2 in $Q$, and we may assume that $a_{11,3}=v$ 
and $a_{12,3}=a_{13,3}=a_{14,3}=v^\perp$. 
It follows easily that $a_{11,4}=z$ and 
$a_{15,4}=a_{14,4}^\perp=z$. Hence, $\pi_3=4,3,1$ and 
$\pi_4=6,1^2$. This $A$ belongs also to the switching class 12.

Suppose that $a_{1,3}=y^\perp$. 

For convenience set $u:=a_{1,4}$.
Since $a_{2,3}=a_{1,3}=y^\perp$ and $P\in\cO(2)$, by permuting the rows 3,4,5 of $A$, we can assume that $a_{3,3}=y$.
The variable $y$ may occur in $P$ once or twice. 
We distinguish these two possibilities.

Assume that $y$ occurs only once in $P$. Since $y$ has multiplicity 3 in $M_0$, we may assume that 
$a_{7,3}=a_{6,3}^\perp=y$. Since the rows 1 and 6 are orthogonal, we have $a_{6,4}=u^\perp$. Since the rows 7 and 8 are orthogonal, we have $a_{7,4}=z^\perp$. Moreover, Proposition \ref{pp:sused} (iii) applied to $M_0$ and its submatrix $A[7,8;4]$ implies that 
$z$ occurs only once in $M_0$. Thus, $u$ and $z$ are independent. 
Note that $a_{4,3}$ occurs only once in $M_0$ and that   $a_{5,3}=a_{4,3}^\perp$. This implies that $a_{4,4}=a_{5,4}$. 
As $u$ has multiplicity 3 in $M_0$, it follows that $a_{3,4}=u$ and $a_{4,4}=a_{5,4}=u^\perp$. 
As $A$ is maximal, by Proposition \ref{pp:sused} (iii), we have $\mu(a_{4,3})=1$. 

There are three possibilities for the variable $v$ as follows: 
$v=y$, $v=y^\perp$ and $v,y$ are independent. 

First, let $v=y$. Since rows 8 and 9 are orthogonal, we have 
$w=z^\perp$, i.e.,  $a_{9,4}=a_{10,4}=z^\perp$.  
Since $y$ has multiplicity 3 in $M_2$, we must have 
$a_{14,3}=a_{15,3}=y^\perp$ and $a_{14,4}=a_{15,4}^\perp$. 
By permuting the rows 11,12,13 of $A$, we may assume that $a_{11,3}=y^\perp$. 
Hence, the entry $a_{12,3}$ occurs only once in $M_2$ and 
$a_{13,3}=a_{12,3}^\perp$. This implies that $a_{12,4}=a_{13,4}$. 
As $A$ is maximal, by Proposition \ref{pp:sused} (iii), we have $\mu(a_{12,3})=\mu(a_{14,4})=1$. 
Now one can verify that $A$ belongs to the switching class 12.

Second, let $v=y^\perp$. At least one of the entries $a_{14,3},a_{15,3}$ is not equal to $y^\perp$. 
By interchanging the rows 14 and 15 if necessary, we may assume that $a_{15,3}\ne y^\perp$.  As rows 15 and 16 are orthogonal, we must have $a_{15,4}=z^\perp$. We cannot have $a_{15,3}=y$ since 
then by Proposition \ref{pp:sused} (iii) applied to $A$ and its submatrix $A[7,15;1]$ would give a contradiction. 
We infer that $y$ must occur twice in $Q$ and that $\mu(a_{14,3})=1$. By permuting the rows 11,12,13 we may assume that $a_{12,3}=y^\perp$ and $a_{13,3}=a_{14,3}=y$. 
By Proposition \ref{pp:sused} (iii) applied to $A$ and its submatrix $A[12,13;4]$, we deduce that $\mu(a_{12,4})=1$. 
Since rows 14 and 16 are orthogonal, we must have $w=z$.
Now one can verify that $A$ belongs to the switching class 12.

Third, let $v$ and $y$ be independent. Then $v$ has multiplicity 
2 in $Q$ and we may assume that $a_{11,3}=v$, 
$a_{12,3}=a_{13,3}=a_{14,3}=v^\perp$ and $a_{15,3}=y^\perp$. 
Since rows 9 and 16 are orthogonal, we must have $w=z^\perp$.
Now one can verify that $A$ belongs to the switching class 13.

Now assume that $y$ occurs twice in $P$. 
By permuting the rows 3,4,5 of $A$ we may assume that 
$a_{3,3}=y^\perp$ and $a_{4,3}=a_{5,3}=y$. 
By inspecting the matrix \eqref{eq:3qubit=8row}, we infer that 
we must have $a_{3,4}=z^\perp$. 
Since the rows 2 and 3 are orthogonal, we must have $u=z$. 
By applying Proposition \ref{pp:sused} (iii) to $A$ and the submatrix  $A[4,5;4]$, we conclude that $a_{4,4}=a_{5,4}^\perp$ and $\mu(a_{4,4})=1$.  Since $z=u$ has multiplicity 3 in $M_0$, 
we must have $a_{6,4}=a_{7,4}=z^\perp$. 

Since the rows 8 and 9 are orthogonal, $v=y^\perp$ or 
$w=z^\perp$. 
Note that we cannot have $v=y^\perp$ and $w=z^\perp$. 
If $w=z$ then $z$ has multiplicity 3 in $M_2$ and we may assume that $a_{11,4}=z^\perp$. By applying Corollary \ref{cr:ir=1,3} to  the rows 3 and 8 of $M_2$, we obtain that $a_{11,3}=y^\perp$. 
By applying Proposition \ref{pp:sused} (iii) to the submatrix  $A[3,11;1]$, we obtain a contradiction. 
We conclude that $w\ne z$. Similarly, $v\ne y$.

Thus, if $v=y^\perp$ then $w$ and $z$ are independent, 
and $z$ must occur only once in $M_2$. It follows that $z^\perp$
does not occur in $Q$. 
Hence, we may assume that $a_{11,4}=w$, 
$a_{12,4}=a_{13,4}=a_{14,4}=w^\perp$ and $a_{15,4}=z^\perp$.
By applying Proposition \ref{pp:sused} (iii) to $A$ and the submatrix  $A[12,13;3]$, we obtain that 
$a_{12,3}=a_{13,3}^\perp$ and $\mu(a_{13,3})=1$. 
It follows that $a_{14,3}=y^\perp$ and $a_{15,4}=y$. Thus, 
$\pi_3=6,1^2$, $\pi_4=4,3,1$ and one can verify that $A$ belongs to the switching class 12.
Similarly, if $w=z^\perp$ then $v$ and $y$ are independent, 
$\pi_3=4,3,1$, $\pi_4=6,1^2$ and $A$ belongs to the switching class 12.

Case 3: $\pi_1=6,2$. 

By permuting the rows of $A$ and renaming the variables, if necessary, we may assume that 
$[~a~a~a~a~a~a~b~b~a^\perp~a^\perp~a^\perp~a^\perp
~a^\perp~a^\perp~b^\perp~b^\perp~]^T$
is the first column of $A$. 
Assume that $a_{7,j}\ne a_{8,j}$ for at least two indices 
$j\in\{2,3,4\}$. Since both submatrices
$$
M_0=\left[\begin{array}{c} N_0 \\ N_1 \end{array} \right], \quad  
M_3=\left[\begin{array}{c} N_3 \\ N_0 \end{array} \right]
$$ 
belong to $\cO(3)$, Lemma \ref{le:3q-2rows} implies that the 
$2\times3$ submatrices $N_1$ and $N_3$ may differ only in the 
order of rows. By interchanging the last two rows of $A$ 
(if necessary) we may assume that $N_1=N_3$. This switch 
will not change the first column of $A$ because  $a_{15,1}=a_{16,1}=b^\perp$. 
It is now easy to verify that if we replace the entries 
$a_{8,1}=b$ and $a_{16,1}=b^\perp$ in $A$ with new vector 
variables $x$ and $x^\perp$, respectively, then the 
modified matrix $A$ will still belong to $\cO(4)$. 
This gives a contradiction since $A$ is maximal. 
Hence, the equality $a_{7,j}=a_{8,j}$ must hold for exactly
two indices $j\in\{2,3,4\}$. 
We may assume that $c:=a_{7,2}=a_{8,2}$ and
$d:=a_{7,3}=a_{8,3}$. 

Since $M_0,M_3\in\cO(3)$, we must also have
$a_{15,2}=a_{16,2}=c$ and $a_{15,3}=a_{16,3}=d$. 
Since the rows 7 and 8 of $A$ are orthogonal, we must have 
$a_{8,4}=a_{7,4}^\perp$. Similarly, $a_{16,4}=a_{15,4}^\perp$. 
Since $A$ is maximal, by applying 
Proposition \ref{pp:sused} (iii) to the submatrix $A[7,8;4]$, 
we deduce that $\mu(a_{7,4})=1$. Similarly, $\mu(a_{15,4})=1$. 
In particular, the vector variables $a_{7,4}$ and $a_{15,4}$ 
are independent.

Subcase 3a: Some $M_i$, say $M_0$, is reducible. 

Since $\mu(a_{7,4})=1$, either $c$ or $d$, say $c$, must have 
multiplicity 4 in $M_0$. Thus there are four entries $c$ in the 
first column of $M_0$ and at least two entries $c$ in the 
first column of $M_2$. As $\mu(x)\le6$ for all entries $x$ 
of $A$, we conclude that $\mu(c)=6$. Hence, $c$ has multiplicity 2 in $M_2$. By Corollary \ref{cr:ir=1,3}, $M_2$ is reducible. 
By permuting the first six rows of $A$, we may assume that the first column of $M_0$ is
$[~c^\perp~c^\perp~c^\perp~c^\perp~c~c~c~c~]^T$. By permuting the 
rows 9 to 14 of $A$, we may also assume that 
$a_{9,2}=a_{10,2}=c^\perp$.
As $\mu(c^\perp)=6$, we have $a_{i,2}\ne c^\perp$ for 
$i\in\{11,12,13,14\}$. 
Since the row 15 of $A$ is orthogonal to the rows 
5,6,11,12,13,14 and $\mu(a_{15,4})=1$, we deduce that 
$a_{i,3}=d^\perp$ for $i\in\{5,6,11,12,13,14\}$. 
It follows that $a_{5,4}=a_{6,4}^\perp$ and that the multiplicity of $d$ in $M_2$ must be 4. 
Consequently, $a_{9,3}=a_{10,3}=d$, and $a_{9,4}=a_{10,4}^\perp$. 
By applying Proposition \ref{pp:sused} (iii) to the submatrices $A[5,6;4]$ and $A[9,10;4]$, we deduce that 
$\mu(a_{5,4})=\mu(a_{9,4})=1$. 

Note that the submatrices $A[1,2,3,4;3,4]$ and 
$A[11,12,13,14;2,4]$ must belong to $\cO(2)$. 
By using Proposition \ref{pp:sused} (iii-iv) we deduce that these
two submatrices are maximal in $\cO(2)$ and have no vector variable in common. Hence, these submatrices are equivalent to 
the second matrix in \eqref{mat:2-q}.
By applying the switching operations (if necessary) on these 
two submatrices, we may assume that $a_{1,3}$ and $a_{11,2}$ 
have multiplicity 2 in $A[1,2,3,4;3,4]$  and 
$A[11,12,13,14;2,4]$, respectively. Then we have 
$\pi_1=\pi_2=\pi_3=6,2$ and $\pi_4=1^8$. 
Thus $A$ is weakly equivalent to the third matrix in \eqref{eq:sw1-3}, i.e., $A$ belongs to the switching class 3.

Subcase 3b: All $M_i$ are irreducible. 

Thus, each $M_i$ is equivalent to the matrix 
\eqref{eq:3qubit=8row}.
By Corollary \ref{cr:ir=1,3}, both $c$ and $d$ must have multiplicity 3 in both $M_0$ and $M_1$. Hence, $\mu(c)=6$. 
By permuting the rows of $N_0$ and $N_2$, we may assume that 
$[~x~x^\perp~c^\perp~c^\perp~c^\perp~c~c~c~y~y^\perp~c^\perp
~c^\perp~c^\perp~c~c~c~]^T$ is the second column of $A$. 
As $M_0$ and $M_2$ are irreducible and $c$ is independent from 
$x$ and $y$, by inspecting the matrix \eqref{eq:3qubit=8row}, we deduce that $a_{1,3}=a_{2,3}$, $u:=a_{1,4}=a_{2,4}$, 
$a_{9,3}=a_{10,3}$, and $v:=a_{9,4}=a_{10,4}$. 
By Corollary \ref{cr:ir=1,3}, $u$ and $v$ have multiplicity 3 in $M_0$ and $M_2$, respectively. 
As each of the rows 1,2,6,9,10,14 of $A$ is orthogonal to rows 
7 and 8, we infer that $a_{i,3}=d^\perp$ for $i=1,2,6,9,10,14$. 
As rows 1 and 6 of $A$ are orthogonal, we have $a_{6,4}=u^\perp$.
As rows 9 and 14 of $A$ are orthogonal, we have 
$a_{14,4}=v^\perp$.

Exactly one of the entries $a_{i,3}$, $i=3,4,5$ is equal to $d$. 
By permuting the rows 3,4,5 of $A$, we may assume that 
$a_{5,3}=d$. 
Similarly, we may assume that $a_{13,3}=d$.

Since $a_{5,2}=a_{6,2}^\perp$ and 
$a_{5,3}=a_{6,3}^\perp$, by inspection of \eqref{eq:3qubit=8row}, 
we infer that we must also have $a_{5,4}=a_{6,4}^\perp$, i.e., 
$a_{5,4}=u$. Similarly, $a_{13,4}=v$.
It follows that $a_{3,4}=a_{4,4}=u^\perp$ and 
$a_{11,4}=a_{12,4}=v^\perp$.

By applying Proposition \ref{pp:sused} (iii) to the $2\times1$ submatrix $A[1,2;2]$, we deduce that $x$ and $y$ must be independent. Similarly, $a_{3,3}$ and $a_{11,3}$ must be independent, as well as $a_{7,4}$ and $a_{15,4}$. Finally, 
the maximality of $A$ implies that $u$ and $v$ are independent.
Thus $\pi_2=\pi_3=6,1^2$ and $\pi_4=3^2,1^2$, and $A$ belongs to the switching class 10.

Case 4: $\pi_1=6,1^2$. 

We may assume that 
$[~a~a~a~a~a~a~b~c~a^\perp~a^\perp~a^\perp~a^\perp
~a^\perp~a^\perp~b^\perp~c^\perp~]^T$ 
is the first column of $A$. 
Since both $M_1,M_2\in\cO(3)$, Lemma \ref{le:1-row} implies that $N_1=N_4$. Similarly, since $M_2,M_3\in\cO(3)$, we have 
$N_2=N_5$.
Set $x:=a_{7,2}$, $y:=a_{7,3}$ and $z:=a_{7,4}$. 
As $N_1=N_4$ we also have $a_{15,2}=x$, $a_{15,3}=y$ and 
$a_{15,4}=z$. 
As $N_2=N_5$, we have $a_{16,j}=a_{8,j}$ for $j=2,3,4$. 

Subcase 4a: Some $M_i$, say $M_0$, is reducible. 

Since $M_0$ is reducible, at least one of $x,y,z$ must have multiplicity 4 in $M_0$. By permuting the last three columns 
of $A$ (if necessary) we may assume that $x$ has multiplicity 4 in $M_0$. In particular, $a_{8,2}\in\{x,x^\perp\}$. 

Suppose that $a_{8,2}=x^\perp$. Then also $a_{16,2}=x^\perp$. 

Assume that $M_3$ is irreducible. Then Corollary \ref{cr:ir=1,3} and  $\mu(x)\le6$ imply that $x$ must occur only once in $M_3$. 
Since $M_3$ is equivalent to the matrix \eqref{eq:3qubit=8row}, 
we conclude that $a_{16,3}=y$ and $a_{16,4}=z$. It follows 
that also $a_{8,3}=y$ and $a_{8,4}=z$. We can now select a new 
independent variable $r$ and replace the entries $x$ and $x'$ 
with $r$ and $r^\perp$ respectively, but only in the four positions $a_{i,2}$, $i=7,8,15,16$. We obtain a new matrix in $\cO(4)$ showing that $A$ is not maximal. This contradicts our hypothesis. We conclude that $M_3$ must be reducible.

By permuting the first 6 rows of $A$, we may assume that 
$a_{i,2}$ is equal to $x^\perp$ for $i=1,2,3$ and equal to $x$ 
for $i=4,5,6$. As $\mu(x)\le6$, the multiplicity of $x$ in $M_3$ 
is 1 or 2. Hence, either $y$ or $z$ must have multiplicity 4 in $M_3$. By interchanging the last two columns of $A$ (if necessary) we may assume that $y$ has multiplicity 4 in $M_3$. 
As $a_{8,3}=a_{16,3}$, we have $a_{8,3}\in\{y,y^\perp\}$. 
By permuting the rows of $N_3$, we may assume that 
$a_{i,3}$ is equal to $y^\perp$ for $i=9,10,11$, it is equal to 
$y$ for $i=13,14$, and that $a_{12,3}=a_{8,3}^\perp$. 

If $a_{8,3}=y$ then either $x$ or $z$ must have multiplicity 2 
in the submatrix $A[13,14,15,16;2,4]\in\cO(2)$. If $z$ has multiplicity 2, then $z=u^\perp$ and $\mu(z)=6$. We can interchange the two columns of 
$A[9,10,11,12;2,4]\in\cO(2)$ (a switching operation) to obtain 
a matrix in $\cO(4)$ having $\pi_4=6,2$. This contradicts our hypothesis that $A$ is chosen in its switching class to have the largest possible partition $\pi_1$. A similar argument gives 
a contradiction if $x$ has multiplicity 2 in 
$A[13,14,15,16;2,4]$. 

Thus, we may assume that 
$a_{8,3}=y^\perp$, and consequently $a_{16,3}=y^\perp$ and 
$a_{12,3}=y$. Since the submatrix $A[12,13,14,15;2,4]\in\cO(2)$ and $a_{15,2}=x$, the variable $x^\perp$ must also occur in this 
submatrix. By permuting the rows 12,13,14 of $A$, we may assume that $a_{14,2]=x^\perp}$. We infer that $\mu(x^\perp)=6$ and that $r:=a_{12,2}=a_{13,2}^\perp$, where the variable $r$ is independent from $x$. 
Since $x$ has multiplicity 1 in $A[12,13,14,15;2,4]$, 
the entry $a_{15,4}=z$ must have multiplicity 2 in this submatrix. It follows that $a_{14,4}=z$ and 
$a_{12,4}=a_{13,4}=z^\perp$. 
Since $\mu(x)=\mu(x^\perp=6$, by permuting the rows 9,10,11 of 
$A$, we may assume that $a_{9,2}=x$. It follows that 
$s:=a_{10,2}=a_{11,2}^\perp$ where $s$ is independent from $x$. 
Since $x$ has multiplicity 1 in $A[8,9,10,11;2,4]\in\cO(2)$, 
the entry $a_{8,4}=u^\perp$ must have multiplicity 2 in this submatrix. It follows that $a_{9,4}=u^\perp$ and 
$a_{10,4}=a_{11,4}=u$. 
As $\mu(z)\le6$, $u$ and $z$ must be independent. Since $A$ is 
maximal, $r$ and $s$ must be also independent. 
Hence $A$ belongs to the switching class 14.
 
Suppose now that $a_{8,2}=x$. Then also $a_{16,2}=x$. 

We may assume that $a_{i,2}$ is equal to $x$ for $i=5,6$ and 
it is equal to $x^\perp$ for $i=1,2,3,4,9,10$. 
Since $\pi_1\ge\pi_2$, by permuting the last four rows of $N_3$, we may assume that $a_{11,2}=u$, $a_{12,2}=u^\perp$, 
$a_{13,2}=v$, $a_{14,2}=v^\perp$, with $u,v,x$ independent. 

If $a_{8,3}=y$ then also $a_{16,3}=y$ and 
$a_{8,4}=a_{16,4}=z^\perp$. 
It follows easily that $a_{i,3}=y^\perp$ for 
$i=11,12,13,14$. 
By interchanging the two columns of 
$A[11,12,13,14;2,4]$ (a switching operation) we see 
that $A$ is weakly equivalent to a matrix with $\pi_2=6,2$.  
As $\pi_2>\pi_1=6,1^2$, we can discard this $A$.
Similarly, we can discard $A$ if $a_{8,4}=z$. 

Since the rows 7 and 8 of $A$ are orthogonal, by interchanging 
the last two columns of $A$ (if necessary) 
we may assume that $a_{8,4}=a_{16,4}=z^\perp$. 

For convenience set $w:=a_{8,3}$ and recall that $w\ne y$. Note that the submatrix $A[12,14,15,16;3,4]\in\cO(2)$. 
Assume that at least one of $a_{12,4}$ and $a_{14,4}$ is 
independent from $z$, say $a_{12,4}$. Then $a_{12,3}$ has to be 
orthogonal to both $w$ and $y$. This is impossible because 
$w\ne y$. Thus, $z$ must have multiplicity 
2 in $A[12,14,15,16;3,4]$. 
By permuting the rows of $N_3$ (if necessary), we may assume that $a_{12,4}=a_{14,4}^\perp=z$. 
Since the rows 12 and 15 of $A$ are orthogonal, we infer 
that $a_{12,3}=y^\perp$. 
Since the rows 14 and 16 of $A$ are orthogonal, we infer 
that $a_{14,3}=w^\perp$. 
Since also $A[12,13,15,16;3,4]\in\cO(2)$, Lemma \ref{le:1-row} implies that $a_{13,3}=a_{14,3}=w^\perp$ and 
$a_{13,4}=a_{14,4}=z^\perp$. 
Similarly, since $A[11,14,15,16;3,4]\in\cO(2)$ we have  
$a_{11,3}=a_{12,3}=y^\perp$ and 
$a_{11,4}=a_{12,4}=z$. 
Assume that $a_{9,4}$ and $z$ are independent. Then, 
since $A[9,10,11,13;3,4]\in\cO(2)$, the entry $a_{9,3}$ must be orthogonal to both $a_{11,3}=y^\perp$ and $a_{13,3}=w^\perp$. 
As $w\ne y$, this is impossible. Thus, we must have 
$\{a_{9,4},a_{10,4}\}=\{z,z^\perp\}$. 
By interchanging the rows 9 and 10 of $A$ (if necessary), 
we may assume that $a_{9,4}=z$ and $a_{10,4}=z^\perp$. 
Since rows 9 and 11 are orthogonal, we must have $a_{9,3}=y$. 
Since rows 10 and 13 are orthogonal, we must have $a_{10,3}=w$. 
Assume that $z$ occurs twice in $A[5,6,7,8;3,4]\in\cO(2)$. 
Then $\mu(z)=6$ and by interchanging the two columns of 
$A[1,2,3,4;3,4]\in\cO(2)$ (a switching operation), the last 
column of $A$ will have $6,2$ as the associated partition. 
This contradicts our choice of $A$. 
We conclude that $z$ must occur only once in $A[5,6,7,8;3,4]$. 
By interchanging the rows 5 and 6 of $A$ (if necessary), we 
may assume that $a_{5,4}=a_{6,4}^\perp=r$, where $r$ is a 
variable independent from $z$. 
Since the row 6 of $A$ is orthogonal to rows 7 and 8, we 
infer that $a_{5,3}$ is orthogonal to both $w$ and $y$. 
As $w\ne y$ this is impossible.

Subcase 4b: All $M_i$ are irreducible. 

Since the rows 7 and 8 are orthogonal, by permuting the last three columns of $A$, we may assume that $a_{8,2}=x^\perp$. We discuss three cases, namely $a_{8,3}\in\{y,y^\perp\}$ and $a_{8,3}$ and $y$ are independent. 

Suppose that $a_{8,3}=y^\perp$. As $M_0$ is irreducible, 
we must have also $a_{8,4}=z^\perp$. Since $N_5=N_2$ and $M_0$ and $M_3$ are equivalent to \eqref{eq:3qubit=8row}, we infer that 
each of the variables $x,y,z$ has multiplicity 3 in $M_0$ and 
in $M_3$. Hence, by permuting the rows of $N_0$ and those of 
$N_3$, we may assume that 
$u:=a_{1,2}=a_{2,2}^\perp$, 
$v:=a_{3,3}=a_{4,3}^\perp$, 
$w:=a_{5,4}=a_{6,4}^\perp$ 
and
$r:=a_{9,2}=a_{10,2}^\perp$, 
$s:=a_{11,3}=a_{12,3}^\perp$, 
$t:=a_{13,4}=a_{14,4}^\perp.$ 
Since $A$ is maximal, the vector variables  
$u,v,w,r,s,t,x,y,z$ must be independent. 
By interchanging the rows 7 and 8 as well as the rows 
15 and 16, we may assume that 
$a_{3,2}=a_{4,2}=x$ and $a_{5,2}=a_{6,2}=x^\perp$. 
Then it follows that 
$a_{1,3}=a_{2,3}=y^\perp$, $a_{5,3}=a_{6,3}=y$, 
$a_{1,4}=a_{2,4}=z$, $a_{3,4}=a_{4,4}=z^\perp$. 
There are now two possibilities for the block $N_3$.
First, $a_{i+8,j}=a_{i,j}$ for $j=2,3,4$ if $a_{i,j}$ is independent from $u,v,w$. 
In that case $A$ is not maximal as we can replace the entries
$a_{1,1}$ and $a_{2,1}$ with $p$ and the entries 
$a_{9,1}$ and $a_{10,1}$ with $p^\perp$, where $p$ is a new 
independent variable. The new matrix is still in $\cO(4)$, 
which contradicts the maximality of $A$.  
Thus we can discard this possibility. 
Second, $a_{i+8,j}=a_{i,j}^\perp$ for $j=2,3,4$ if $a_{i,j}$ is independent from $u,v,w$. 
In that case $A$ is equivalent to the representative of the
switching class 11, i.e., the second matrix in \eqref{eq:sw7-9}. 

Suppose that $a_{8,3}=y$. As $y$ has multiplicity 3 in  
$M_0$ and $M_3$, we may assume that the third column of $A$ is 
$[~u~u^\perp~y^\perp~y^\perp~y^\perp~y~y~y
~v~v^\perp~y^\perp~y^\perp~y^\perp~y~y~y~]^T$ 
where $u,v,y$ are independent. 
Since $M_0$ and $M_3$ are irreducible, 
we must have $a_{1,2}=a_{2,2}$, $a_{1,4}=a_{2,4}$
$a_{10,2}=a_{9,2}$ and $a_{10,4}=a_{9,4}$. 

Assume that the multiplicity of $x$ in $M_0$ is 1. 
Then we must have $a_{8,4}=z$, $a_{16,4}=z$, and so 
$x$ has also multiplicity 1 in $M_3$. 
By interchanging the two columns of the submatrix 
$A[7,8,15,16;1,2]\in\cO(2)$ (a switching operation), 
we obtain a matrix in $\cO(4)$ with $6,2$ as the partition 
associated to the first column. Hence, we can discard 
this possibility.

We conclude that the multiplicity of $x$ in $M_0$ and 
in $M_3$ is 3. 
By inspection of the matrix \eqref{eq:3qubit=8row}, we 
conclude that $p:=a_{8,4}$ and $z$ are independent. 
Since rows 1 and 6 are orthogonal to rows 7 and 8, we must have 
either 
\bea \label{eq:alt0-1}
a_{1,2}=x,~ a_{1,4}=z^\perp,~ 
a_{6,2}=x^\perp,~ a_{6,4}=p^\perp
\eea
or
\bea \label{eq:alt0-2}
a_{1,2}=x^\perp,~ a_{1,4}=p^\perp,~ 
a_{6,2}=x,~ a_{6,4}=z^\perp.
\eea
The same two alternatives apply to the corresponding entries 
of $M_3$, namely either
\bea \label{eq:alt3-1}
a_{9,2}=x~, a_{9,4}=z^\perp,~ 
a_{14,2}=x^\perp,~ a_{14,4}=p^\perp
\eea
or 
\bea \label{eq:alt3-2}
a_{9,2}=x^\perp,~ a_{9,4}=p^\perp,~ 
a_{14,2}=x,~ a_{14,4}=z^\perp.
\eea
If the first alternative holds in both $M_0$ and $M_3$, then 
$A$ is not maximal since we can replace the entries 
$a_{6,1}$ and $a_{14,1}$ with a new vector variable and its 
perpendicular. Thus the first alternative cannot hold in 
both $M_0$ and $M_3$. Similarly, the second alternative 
cannot hold in both $M_0$ and $M_3$. If different alternatives 
hold in $M_0$ and $M_3$ then one can verify that $A$ belongs 
to the switching class 14. 
For instance, assume that \eqref{eq:alt0-1} and 
\eqref{eq:alt3-2} hold. 
Note that $a_{2,2}=a_{1,2}=x$ and $a_{2,4}=a_{1,4}=z^\perp$. 
By permuting the rows 3,4,5 of $N_0$, we may assume that 
$a_{3,2}=x^\perp$. Since the first two entries of row 3 of $N_0$ are $x^\perp$ and $y^\perp$ and $N_1=[~x~y~z~]$, we infer that 
$a_{3,4}=z^\perp$ and $a_{4,4}=a_{5,4}=z$. 
Similarly, we may assume that $a_{11,2}=x$ and obtain that 
$a_{11,4}=p^\perp$ and $a_{12,4}=a_{13,4}=p$. 
Since $A$ is maximal, the variables $a_{4,2},a_{12,2},x$ must 
be independent. One can now verify that $A$ is equivalent 
to the second matrix in \eqref{eq:sw13-15}.

Finally, suppose that $u:=a_{8,3}$ and $y$ are independent. 
Since we have already handled the cases 
$a_{8,3}\in\{y,y^\perp\}$, we may assume that 
$v:=a_{8,4}$ and $z$ are independent.
It follows that $x$ must have multiplicity 3 in $M_0$ and $M_3$. 
We may assume that the second coulmn of $A$ is  
$[~w~w^\perp~x^\perp~x^\perp~x~x~x~x^\perp~
d~d^\perp~x^\perp~x^\perp~x~x~x~x^\perp~]^T$,
where $d,w,x$ are independent variables. 
Since $A[1,5,6,7;3,4]\in\cO(2)$, $y$ or $z$ must have 
multiplicity 2 in this submatrix, and multiplicity 3 in $M_0$ since $M_0$ is irreducible. We may assume that $y$ has 
multiplicity 3 in $M_0$ and $u$ multiplicity 1. 
As $a_{1,3}=a_{2,3}$, $a_{1,4}=a_{2,4}$ and 
$A[2,3,4,8;3,4]\in\cO(2)$, we may assume that $a_{4,3}=u^\perp$. 
Since row 8 of $A$ is orthogonal to the first three rows, 
we deduce that $a_{1,4}=a_{2,4}=a_{3,4}=v^\perp$. 
It follows that $a_{4,4}=v$, and we may assume that 
$a_{5,4}=z^\perp$ and $a_{6,4}=v$. 
Since the row 7 of $A$ is orthogonal to the first two rows, 
we deduce that $a_{1,3}=a_{2,3}=y^\perp$. 
Hence, $a_{3,3}=a_{5,3}=y$. 
As $A[9,13,14,15;3,4]\in\cO(2)$, $y$ or $z$ must have 
multiplicity 2 in this submatrix, and so multiplicity 3 
in $M_3$.
As also $A[10,13,14,15;3,4]\in\cO(2)$, we deduce that 
$a_{9,3}=a_{10,3}$ and $a_{9,4}=a_{10,4}$. 
If $y$ has multiplicity 2 in $A[9,13,14,15;3,4]$, then
$a_{9,3}=a_{10,3}=y$ and $a_{13,3}=a_{14,3}=y^\perp$. 
We may assume that $a_{11,3}=u^\perp$ and  $a_{12,3}=y^\perp$.
Thus $\pi_3=6,2>\pi_1$ and we have a contradiction. 
We conclude that $z$ must have multiplicity 2 in $A[9,13,14,15;3,4]$. Then we must have 
$a_{9,4}=a_{10,4}=z^\perp$, and we may assume that 
$a_{14,4}=a_{13,4}^\perp=z$ and that $a_{11,4}=z$, 
$a_{12,4}=v^\perp$. 
Since rows 14 and 15 are orthogonal, we have $a_{14,3}=y^\perp$.  
Since rows 11 and 16 are orthogonal, we have $a_{11,3}=u^\perp$.  
Since rows 8 and 9 are orthogonal, we have 
$a_{9,3}=a_{10,3}=u^\perp$. Consequently, $a_{12,3}=a_{13,3}=u$.
Thus $\pi_3=\pi_4=4^2$ and $A$ is in switching class 15.

Thus we have shown that there are exactly 15 switching classes 
in $\cO(4)$ and we have obtained the list of their representatives as given in the theorem. To complete the proof, one has to apply all possible switching operations to these 15 representatives and select the nonequivalent matrices among them. In each case there are just a few such operations. The representatives of the switching classes 4,11,12,13,14,15 admit no nontrivial switching operations. For the other switching classes we list their equivalence classes in the appendix. We omit the details.
\epf

\section{Applications}

In this section we explain the mathematical and physical meaning and application of our results. 

\subsection{Construction of OPBs in higher dimensions}
\label{subsec:phy}


We use the 4-qubit OPBs to construct reducible 5-qubit OPBs as follows. If 
$\ket{\a_1},\ldots\ket{\a_{16}}$ and 
$\ket{\b_1},\ldots\ket{\b_{16}}$ are two 4-qubit OPBs, then 
$\ket{0,\a_1},\ldots\ket{0,\a_{16}}$ and $\ket{1,\b_1},\ldots\ket{1,\b_{16}}$ are two reducible 5-qubit OPBs. 
Since we have classified all 4-qubit OPBs in Theorem 
\ref{thm:4qubit}, this construction covers all reducible 5-qubit OPBs. 
By using the same idea, we can construct all reducible 
$(n+1)$-qubit OPBs provided that all $n$-qubit OPBs are known. 

We can construct OPBs by using the tensor product of two OPBs. Let $\cH'=\cH'_1\ox\cdots\ox\cH'_n$ be another $n$-partite Hilbert space 
with $\dim \cH'_j = d'_j$ and $D'=d'_1\cdots d'_n$.
Let 
$\cA'=\{\ket{a'_{j,1},\ldots,a'_{j,n}},~~~j=1,\ldots,D'\}\in\cH'$ be an OPB. 
Then 
\bea
\cA\ox\cA':=\{\ket{a_{i,1},a'_{j,1}}\ox\cdots\ox\ket{a_{i,n},a'_{j,n}},~~~i=1,\cdots,D,~~j=1,\cdots,D'\}
\eea 
is an OPB of the $n$-partite Hilbert space 
$\ox_{i=1}^n \left(\cH_i\ox\cH'_i\right)$. We have

\bl
\label{le:reXany}
One of $\cA$ and $\cA'$ is reducible if and only if $\cA\ox\cA'$ is reducible.
\el
\bpf
The ``only if'' part is trivial. Let us prove the ``if'' part. Suppose $\cA\ox\cA'$ is reducible. Without any loss of generality, 
we may assume that there exists a nontrivial partition $(P,Q)$ of 
the set $\{1,\ldots,D\}\times\{1,\ldots,D'\}$ such that 
$\ket{a_{j,1},a'_{k,1}}\perp\ket{a_{j',1},a'_{k',1}}$ for all 
$(j,k)\in P$ and $(j',k')\in Q$. Let us set 
$P_j=\{k:(j,k)\in P\}$ and $Q_j=\{k:(j,k)\in Q\}$. 
If both $P_j$ and $Q_j$ are nonempty for some $j$, then 
$\cA'$ is reducible. Otherwise, $\cA$ is reducible. 
\epf

\subsection{Weak equivalence and controlled unitary operations}
\label{subsec:we}

We have introduced the weak equivalence for maximal 
matrices in $\cO(n)$  in Sec. \ref{sec:weakequiv}. In this subsection we explain, from the viewpoint of practical implementation, why we chose the weak equivalence as the classification criterion.
For example the two matrices in \eqref{eq:3qubit} are weakly equivalent.
Here each pair, say $\{a,a^\perp\}$, represents a qubit o. n. basis. So the two matrices in \eqref{eq:3qubit} represent two 
families of OPBs. 
The second of these matrices is obtained from the first by 
interchanging the two columns of the lower right $4\times2$ submatrix. So we can convert one family to the other by the controlled unitary operation $U=\proj{a}\ox I_4 + \proj{a^\perp}\ox S_2$, where $S_2$ is the SWAP gate on two-qubit state. That is, if $\ket{\a_j}$ is a product state in the first matrix of \eqref{eq:3qubit} , then $U\ket{\a_j}$ is a product state in the second matrix of \eqref{eq:3qubit}. In general, the definition of weak equivalence implies that two weakly equivalent states are convertible by a series of controlled unitaries consisting of an identity and a SWAP gate on certain qubits. The controlled unitaries can be physically implemented with a high probability and accuracy. They have been extensively investigated in recent years \cite{cy14,cy14ap,cy15}. In this sense, one may experimentally implement the conversion of different OPBs using controlled unitaries. This is beneficial to quantum error correction and state preparation. 

We give the formal definition as follows. A bipartite unitary gate $U$ is a \textit{controlled unitary gate} if $U$ is equivalent to $\sum^{d_1}_{j=1}\proj{j}\ox U_j$ or
$\sum^{d_2}_{j=1}V_j \ox \proj{j}$ via local unitaries. We say that $U$ is a controlled unitary from $A$ or $B$ side, respectively.
Furthermore, $U$ is controlled in the computational basis from 
$A$ side if $U=\sum^{d_1}_{j=1}\proj{j}\ox U_j$.

\section{Discussion}
\label{diskusija}

The orthogonal product bases (OPBs) in 
$\cH=\cH_1\otimes\cH_2\otimes\cdots\otimes\cH_n$ 
are easy to describe in the bipartite case $(n=2)$ when, 
say, $d_1=2$. (Recall that $d_i=\dim \cH_i$ and 
$D=\prod d_i$.) 
For this see Proposition \ref{pp:OPB} and 
Corollary \ref{cr:n=d_1=2}. 
However, in general, the bipartite case remains open.

In general, the construction and the classification of OPBs up to local unitary operations reduces to the case of so called irreducible OPBs (see section \ref{sec:opb}). 
The irreducible OPBs have been described and classified in \cite{fs09} in the case of two qutrits $(n=2,~ d_1=d_2=3)$ and the case of three qubits $(n=3,~ d_1=d_2=d_3=2)$. 

The multiqubit case $(d_1=d_2=\cdots=d_n=2)$ is much easier 
to solve than the other cases (apart from those mentioned above). Indeed, we have shown (see Sec. \ref{sec:combinatorial}) that, in the multiqubit case, the construction of OPBs reduces to a purely combinatorial problem.
In the case of four qubits, we were able to solve this 
combinatorial problem. Our main result is that there are 
33 explicit multiparameter families of OPBs of four 
qubits such that any OPB is equivalent to a member of 
one of these families.

We have discussed this combinatorial problem with Vijay 
Ganesh. In his opinion, our combinatorial problem for $n=5,6$ could be solved by using computers. It is an interesting question to discuss the computational complexity of finding the complete characterization of the OPBs of an $n$-qubit system.

Our approach to the problem of construction and classification of OPBs of $n$-qubit system is based on the classification of maximal matrices in $\cO(n)$. As mentioned in section \ref{sec:combinatorial}, to a given OPB 
$\cA:=\{\ket{a_s}=\ket{a_{s,1},\ldots,a_{s,n}}:s=1,\ldots,D\}$ 
we can associate a matrix say $A\in\cO(n)$ simply by setting 
$A[s,j]$ to be a vector variable subject to the following 
conditions:

(i) $A[s,j]=A[t,j]$ if and only if $\ket{a_{s,j}}=\ket{a_{t,j}}$; 

(ii) $A[s,j]=A[t,j]^\perp$ if and only if 
$\ket{a_{s,j}}=\ket{a_{t,j}}^\perp$; 

(iii) a vector variable cannot occur in two different columns 
of the matrix $A$.

It is immediate from this definition that $\cA\in\cF_A$. 
The matrix $A$ does not have to be maximal and is not unique as we can choose the names of vector variables in many ways. However, for two equivalent OPBs their associated matrices in 
$\cO(n)$ will be always equivalent. Thus we obtain a map from 
the set of equivalence classes of OPBs to the set of equivalence 
classes of matrices in $\cO(n)$.

If $A$ is not maximal then there exists $B\in\cO(n)$ such that
$A<B$. In that case we have $\cF_A\subset\cF_B$. Hence the 
relation $\cA\in\cF_M$, with $M\in\cO(n)$, does not determine the equivalence class $[M]$ uniquely. This can be corrected by 
introducing strict families. For $M\in\cO(n)$ we define the 
{\em strict family} $\cF_M^\#$ by setting
$$
\cF_M^\#:=\cF_M\setminus\cup_{N: N<M} \cF_N.
$$

Then, going back to our OPB $\cA$ and its associated matrix 
$A\in\cO(n)$, we have $\cA\in\cF_A^\#$. Moreover, the relation 
$\cA\in\cF_M^\#$ implies that $\cF_M^\#=\cF_A^\#$, i.e., 
$\cA$ belongs to a unique strict family, namely $\cF_A^\#$.

\section*{Acknowledgements}

We thank the two anonymous referees for their helpful comments 
and suggestions. Thanks to them, the paper was greatly improved.
We also thank Vijay Ganesh for the discussion.
LC was supported by Beijing Natural Science Foundation (4173076), the NNSF of China (Grant No. 11501024), and the Fundamental Research Funds for the Central Universities (Grant Nos. 29816133 and 74026601). 
The second author was supported in part by the National Sciences and Engineering Research Council (NSERC) of Canada Discovery 
Grant 5285. 


\section*{Appendix} 
\label{sec:app}

There are 15 switching classes in $\cO(4)$. Their 
representatives are listed in \eqref{eq:sw1-3}-\eqref{eq:sw13-15}. 
On the other hand there are 33 equivalence classes of maximal matrices in $\cO(4)$. 
Each of the six switching classes 4,11,12,13,14,15  contains a  single equivalence class of maximal matrices. 
For each of the remaining nine switching classes, we list below the representatives $A$ of the equivalence classes of maximal 
matrices contained in them. We also record the number $\nu_A$ of 
independent variables that occur in the matrix $A$. 
This number is constant over each switching class. 

Switching class 1: $\nu_A=15$. This switching class is the disjoint union of six equivalence classes. 
\bea
\label{eq:prek-1a}
&& \left[ \begin{array}{cccc}
0 & 0 & 0 & * \\
0 & 0 & 1 & * \\
0 & 1 & c & * \\
0 & 1 & c^\perp & * \\
1 & b & x & * \\
1 & b & x^\perp & * \\
1 & b^\perp & y & * \\
1 & b^\perp & y^\perp & * \\  
\end{array} \right]
\begin{array}{l}8 \\ 4^2 \\ 2^4 \\ 1^8 \\
\end{array}\quad
\left[ \begin{array}{cccc}
0 & 0 & * & 0 \\ 0 & 0 & * & 1 \\
0 & 1 & 0 & * \\
0 & 1 & 1 & * \\
1 & b & c & * \\
1 & b & c^\perp & * \\
1 & b^\perp & x & * \\
1 & b^\perp & x^\perp & * \\
\end{array} \right]
\begin{array}{l}8 \\ 4^2 \\ 2^3,1^2 \\ 2,1^6 \end{array}\quad
\left[ \begin{array}{cccc}
0 & 0 & 0 & * \\
0 & 0 & 1 & * \\
0 & 1 & * & 0 \\
0 & 1 & * & 1 \\
1 & b & * & d \\
1 & b & * & d^\perp \\
1 & b^\perp & c & * \\
1 & b^\perp & c^\perp & * \\
\end{array} \right]
\begin{array}{l}8 \\ 4^2  \\ 2^2,1^4 \\ 2^2,1^4 \end{array}\quad
\\  \label{eq:prek-1b}
&& \left[ \begin{array}{cccc}
0 & 0 & 0 & * \\
0 & 0 & 1 & * \\
0 & 1 & c & * \\
0 & 1 & c^\perp & * \\
1 & b & * & 0 \\
1 & b & * & 1 \\
1 & b^\perp & * & d \\
1 & b^\perp & * & d^\perp \\
\end{array} \right]
\begin{array}{l}8 \\ 4^2 \\ 2^2,1^4 \\ 2^2,1^4 \end{array}\quad
\left[\begin{array}{cccc}
0 & 0 & * & 0 \\
0 & 0 & * & 1 \\
0 & 1 & c & * \\
0 & 1 & c^\perp & * \\
1 & b & 0 & * \\
1 & b^\perp & 0 & * \\
1 & x & 1 & * \\
1 & x^\perp & 1 & * \\
\end{array} \right]
\begin{array}{l}8 \\ 4,2^2 \\ 4,2,1^2 \\ 2,1^6 \end{array}\quad
\left[ \begin{array}{cccc}
0 & 0 & * & 0 \\
0 & 0 & * & 1 \\
0 & 1 & * & d \\
0 & 1 & * & d^\perp \\
1 & b & 0 & * \\
1 & b^\perp & 0 & * \\
1 & x & 1 & * \\
1 & x^\perp & 1 & * \\
\end{array} \right]
\begin{array}{l}8 \\ 4,2^2 \\ 4,1^4 \\ 2^2,1^4 \end{array}\quad
\eea

Switching class 2: $\nu_A=14$. This switching class is the disjoint union of two equivalence classes. 
\bea
\label{eq:prek-2}
&& \left[ \begin{array}{cccc}0 & 0 & c & * \\
0 & 0 & c^\perp & * \\
0 & 1 & x & * \\
0 & 1 & x^\perp & * \\
1 & b & 0 & 0 \\
1 & * & 1 & 0 \\
1 & b & * & 1 \\
1 & b^\perp & 0 & * \\
1 & b^\perp & 1 & 1 \\
\end{array} \right]
\begin{array}{l} 8 \\ 4,3,1 \\ 3,2^2,1 \\ 3,1^5 \\
\end{array}\quad
\left[ \begin{array}{cccc}
0 & 0 & * & d \\
0 & 0 & * & d^\perp \\
0 & 1 & c & * \\
0 & 1 & c^\perp & * \\
1 & * & 1 & 0 \\
1 & b & 0 & 0 \\
1 & b & * & 1 \\
1 & b^\perp & 0 & * \\
1 & b^\perp & 1 & 1 \\
\end{array} \right]
\begin{array}{l} 8 \\ 4,3,1  \\ 3,2,1^3 \\ 3,2,1^3 \end{array}\quad
\eea

Switching class 3: $\nu_A=14$. This switching class is the disjoint union of four equivalence classes. 
\bea
\label{eq:prek-3a}
&& \left[ \begin{array}{cccc}
0 & 0 & 0 & * \\
0 & b & 1 & * \\
0 & b^\perp & 1 & * \\
1 & 0 & c & * \\
1 & 0 & c^\perp & * \\
1 & 1 & 1 & * \\
a & 1 & 0 & * \\
a^\perp & 1 & 0 & * \\
\end{array} \right]
\begin{array}{l} 6,2 \\ 6,2 \\ 6,2 \\ 1^8 \end{array}\quad
\left[ \begin{array}{cccc}
0 & 0 & 0 & * \\
0 & b & 1 & * \\
0 & b^\perp & 1 & * \\
1 & 0 & * & 0 \\
1 & 0 & * & 1 \\
1 & 1 & 1 & * \\
a & 1 & 0 & * \\
a^\perp & 1 & 0 & * \\
\end{array} \right]
\begin{array}{l} 6,2 \\ 6,2 \\ 6,1^2 \\ 2,1^6 \end{array}\quad
\\  \label{eq:prek-3b}
&& \left[ \begin{array}{cccc}
0 & 0 & 0 & * \\
0 & * & 1 & 0 \\
0 & * & 1 & 1 \\
1 & 0 & * & d \\
1 & 0 & * & d^\perp \\
1 & 1 & 1 & * \\
a & 1 & 0 & * \\
a^\perp & 1 & 0 & * \\
\end{array} \right]
\begin{array}{l} 6,2 \\ 6,1^2 \\ 6,1^2 \\ 2^2,1^4 \end{array}\quad
\left[ \begin{array}{cccc}
0 & 0 & 0 & * \\
0 & * & 1 & 0 \\
0 & * & 1 & 1 \\
1 & 0 & * & d \\
1 & 0 & * & d^\perp \\
1 & 1 & 1 & * \\
* & 1 & 0 & x \\
* & 1 & 0 & x^\perp \\
\end{array} \right]
\begin{array}{l} 6,1^2 \\ 6,1^2 \\ 6,1^2 \\ 2^3,1^2 \end{array}\quad
\eea

Switching class 5: $\nu_A=13$. This switching class is the disjoint union of four equivalence classes. 
\bea
\label{eq:prek-5a}
&& \left[ \begin{array}{cccc}
0 & 0 & 0 & * \\
0 & 1 & 0 & 1 \\
0 & b & 1 & * \\
0 & b^\perp & 1 & * \\
1 & 1 & 1 & 0 \\
1 & 1 & * & 1 \\
1 & 0 & c & * \\
1 & 0 & c^\perp & * \\
* & 1 & 0 & 0 \\
\end{array} \right]
\begin{array}{l} 7,1 \\ 6,2 \\ 5,2,1 \\ 3,1^5 \\
\end{array}\quad
\left[ \begin{array}{cccc}
0 & 0 & 0 & * \\
0 & 1 & 0 & 1 \\
0 & b & 1 & * \\
0 & b^\perp & 1 & * \\
1 & 1 & 1 & 0 \\
1 & 1 & * & 1 \\
1 & 0 & * & d \\
1 & 0 & * & d^\perp\\
* & 1 & 0 & 0 \\
\end{array} \right]
\begin{array}{l} 7,1 \\ 6,2 \\ 5,1^3 \\ 3,2,1^3 \end{array}\quad
\\  \label{eq:prek-5b}
&& \left[ \begin{array}{cccc}
0 & 0 & 0 & * \\
0 & 1 & 0 & 1 \\
0 & * & 1 & d \\
0 & * & 1 & d^\perp \\
1 & 1 & 1 & 0 \\
1 & 1 & * & 1 \\
1 & 0 & c & * \\
1 & 0 & c^\perp & * \\
* & 1 & 0 & 0 \\
\end{array} \right]
\begin{array}{l} 7,1 \\ 6,1^2 \\ 5,2,1 \\ 3,2,1^3 \end{array}\quad
\left[ \begin{array}{cccc}
0 & 0 & 0 & * \\
0 & 1 & 0 & 1 \\
0 & * & 1 & d \\
0 & * & 1 & d^\perp \\
1 & 1 & 1 & 0 \\
1 & 1 & * & 1 \\
1 & 0 & * & x \\
1 & 0 & * & x^\perp \\
* & 1 & 0 & 0 \\
\end{array} \right]
\begin{array}{l} 7,1 \\ 6,1^2 \\ 5,1^3 \\ 3,2^2,1 \end{array}\quad
\eea

Switching class 6: $\nu_A=13$. This switching class is the disjoint union of three equivalence classes. 
\bea
\label{eq:prek-6}
&& \left[ \begin{array}{cccc}
0 & 0 & 0 & 0 \\
0 & * & 0 & 1 \\
0 & b & 1 & * \\
0 & b^\perp & 1 & * \\
1 & 1 & 1 & 0 \\
1 & 1 & * & 1 \\
1 & 0 & c & * \\
1 & 0 & c^\perp & * \\
* & 1 & 0 & 0 \\
\end{array} \right]
\begin{array}{l} 7,1 \\ 5,2,1 \\ 5,2,1 \\ 4,1^4 \end{array}\quad
\left[ \begin{array}{cccc}
0 & 0 & 0 & 0 \\
0 & 0 & * & 1 \\
0 & 1 & * & d \\
0 & 1 & * & d^\perp \\
1 & 1 & 1 & 0 \\
1 & * & 1 & 1 \\
1 & b & 0 & * \\
1 & b^\perp & 0 & * \\
* & 0 & 1 & 0 \\
\end{array} \right]
\begin{array}{l} 7,1 \\ 5,2,1 \\ 5,1^3 \\ 4,2,1^2 \end{array}\quad
\left[ \begin{array}{cccc}
0 & 0 & 0 & 0 \\
0 & * & 0 & 1 \\
0 & * & 1 & d \\
0 & * & 1 & d^\perp \\
1 & 1 & 1 & 0 \\
1 & 1 & * & 1 \\
1 & 0 & * & x \\
1 & 0 & * & x^\perp \\
* & 1 & 0 & 0 \\
\end{array} \right]
\begin{array}{l} 7,1 \\ 5,1^3 \\ 5,1^3 \\ 4,2^2 \end{array}\quad
\eea

Switching class 7: $\nu_A=12$. This switching class is the disjoint union of two equivalence classes. 
\bea
\label{eq:prek-7}
&& \left[ \begin{array}{cccc}
0 & 0 & 0 & 0 \\
0 & 0 & 1 & * \\
0 & 1 & c & * \\
0 & 1 & c^\perp & * \\
1 & 1 & 1 & 0 \\
1 & 1 & 0 & * \\
1 & 0 & * & 0 \\
1 & * & 1 & 1 \\
* & 0 & 0 & 1 \\
\end{array} \right]
\begin{array}{l} 7,1 \\ 7,1 \\ 5,2,1 \\ 4,1^4 \end{array}\quad
\left[ \begin{array}{cccc}
0 & 0 & 0 & 0 \\
0 & 0 & 1 & * \\
0 & 1 & * & d \\
0 & 1 & * & d^\perp \\
1 & 1 & 1 & 0 \\
1 & 1 & 0 & * \\
1 & 0 & * & 0 \\
1 & * & 1 & 1 \\
* & 0 & 0 & 1 \\
\end{array} \right]
\begin{array}{l} 7,1 \\ 7,1 \\ 5,1^3 \\ 4,2,1^2 \end{array}\quad
\eea

Switching class 8: $\nu_A=12$. This switching class is the disjoint union of two equivalence classes. 
\bea
\label{eq:prek-8}
&& \left[ \begin{array}{cccc}
0 & 0 & 0 & 0 \\
0 & 1 & 0 & d \\
0 & * & 1 & d \\
0 & 1 & * & d^\perp \\
0 & 0 & 1 & d^\perp \\
1 & 1 & c & * \\
1 & 1 & c^\perp & * \\
1 & 0 & 1 & 1 \\
1 & 0 & * & 0 \\
* & 0 & 0 & 1 \\
\end{array} \right]
\begin{array}{l} 7,1 \\ 7,1 \\ 4,2,1^2 \\ 3^2,1^2 \end{array}\quad
\left[ \begin{array}{cccc}
0 & 0 & 0 & 0 \\
0 & 1 & 0 & d \\
0 & * & 1 & d \\
0 & 1 & * & d^\perp \\
0 & 0 & 1 & d^\perp \\
1 & 1 & * & x \\
1 & 1 & * & x^\perp \\
1 & 0 & 1 & 1 \\
1 & 0 & * & 0 \\
* & 0 & 0 & 1 \\
\end{array} \right]
\begin{array}{l} 7,1 \\ 7,1 \\ 4,1^4 \\ 3^2,2 \end{array}\quad
\eea

Switching class 9: $\nu_A=12$. This switching class is the disjoint union of two equivalence classes. 
\bea
\label{eq:prek-9}
&& \left[ \begin{array}{cccc}
0 & 0 & 0 & 0 \\
0 & b & 0 & 1 \\
0 & b & 1 & * \\
0 & b^\perp & 1 & 0 \\
0 & b^\perp & * & 1 \\
1 & 1 & 0 & 1 \\
1 & 1 & 1 & * \\
1 & 0 & d & * \\
1 & 0 & d^\perp & * \\
* & 1 & 0 & 0 \\
\end{array} \right]
\begin{array}{l} 7,1 \\ 5,3 \\ 5,2,1 \\ 4,1^4 \end{array}\quad
\left[ \begin{array}{cccc}
0 & 0 & 0 & 0 \\
0 & b & 0 & 1 \\
0 & b & 1 & * \\
0 & b^\perp & 1 & 0 \\
0 & b^\perp & * & 1 \\
1 & 1 & 0 & 1 \\
1 & 1 & 1 & * \\
1 & 0 & * & d \\
1 & 0 & * & d^\perp \\
* & 1 & 0 & 0 \\
\end{array} \right]
\begin{array}{l} 7,1 \\ 5,3 \\ 5,1^3 \\ 4,2,1^2 \end{array}\quad
\eea

Switching class 10: $\nu_A=12$. This switching class is the disjoint union of two equivalence classes. 
\bea
\label{eq:prek-10}
&& \left[ \begin{array}{cccc}
0 & 0 & 0 & 0 \\
0 & 1 & 1 & 1 \\
0 & * & 1 & 0 \\
0 & 0 & * & 1 \\
1 & 1 & 1 & d \\
1 & 0 & 0 & d^\perp \\
1 & * & 1 & d^\perp \\
1 & 0 & * & d \\
a & 1 & 0 & * \\
a^\perp & 1 & 0 & * \\
\end{array} \right]
\begin{array}{l} 6,2 \\ 6,1^2 \\ 6,1^2 \\ 3^2,1^2 \\
\end{array}\quad
\left[ \begin{array}{cccc}
0 & 0 & 0 & 0 \\
0 & 1 & 1 & 1 \\
0 & * & 1 & 0 \\
0 & 0 & * & 1 \\
1 & 1 & 1 & d \\
1 & 0 & 0 & d^\perp \\
1 & * & 1 & d^\perp \\
1 & 0 & * & d \\
* & 1 & 0 & x \\
* & 1 & 0 & x^\perp \\
\end{array} \right] 
\begin{array}{l} 6,1^2 \\ 6,1^2 \\ 6,1^2 \\ 3^2,2 \end{array}
\quad
\eea

\end{document}